\documentclass[useAMS,usenatbib]{mn2e}
\usepackage{graphicx}
\usepackage{rotating}
\usepackage{amssymb}

\title[SNe 2013am and 2013K]{SNe 2013K and 2013am: observed and physical properties of two slow, normal Type IIP events}
\author[L. Tomasella et al.]
  {L.~Tomasella,$^1$ E.~Cappellaro,$^1$  M.L. Pumo,$^{1,2,3}$ A. Jerkstrand,$^4$ S.~Benetti,$^1$
  \newauthor N.~Elias-Rosa,$^1$ M.~Fraser,$^5$ C. Inserra,$^{6,7}$ A.~Pastorello,$^1$ M.~Turatto,$^1$  J.P.~Anderson,$^8$
  \newauthor L. Galbany,$^{9}$ C.P. Guti\'errez,$^{7}$ E.~Kankare,$^{6}$ G. Pignata,$^{10,11}$ G.~Terreran,$^{1,12}$ S.~Valenti,$^{13}$ 
  \newauthor C.~Barbarino,$^{14}$ F.E.~Bauer,$^{11,15,16}$ M.T.~Botticella,$^{17}$ T.-W.Chen,$^{18}$ A.~Gal-Yam,$^{19}$
  \newauthor A.~Harutyunyan,$^{20}$ D.A.~Howell,$^{21,22}$ K. Maguire,$^{6}$ A.~Morales Garoffolo,$^{23}$ P.~Ochner,$^1$
  \newauthor S.J.~Smartt,$^6$ S.~Schulze,$^{19}$ D.R.~Young,$^{6}$ L.~Zampieri,$^1$\\
  $^1$INAF Osservatorio Astronomico di Padova, Vicolo dell'Osservatorio 5, 35122 Padova, Italy\\
  $^2$Universit\'a degli Studi di Catania, DIEEI and DFA,  Via Santa Sofia 64, 95123 Catania, Italy  \\
  $^3$INFN Laboratori Nazionali del Sud, Via Santa Sofia 62, 95123 Catania, Italy \\
  $^4$Max-Planck Institut f\"ur Astrophysik, Karl-Schwarzschild-Str. 1, D-85748 Garching, Germany\\
  $^5$School of Physics, O'Brien Centre for Science North, University College Dublin, Belfield, Dublin 4, Ireland\\
  $^6$Astrophysics Research Centre, School of Mathematics and Physics, Queens University Belfast, Belfast BT7 1NN, UK\\
  $^7$Department of Physics \& Astronomy, University of Southampton, Southampton, Hampshire, SO17 1BJ, UK  \\
  $^8$European Southern Observatory, Alonso de C\'ordova 3107, Casilla 19, Santiago, Chile\\ 
  $^9$PITT PACC, Department of Physics and Astronomy, University of Pittsburgh, Pittsburgh, PA 15260, USA\\  
  $^{10}$Departamento de Ciencias Fisicas, Universidad Andres Bello, Avda. Republica 252, Santiago, Chile\\
  $^{11}$Millennium Institute of Astrophysics (MAS), Nuncio Monse\~nor S\'otero Sanz 100, Providencia, Santiago, Chile \\
  $^{12}$Center for Interdisciplinary Exploration and Research in Astrophysics CIERA, Department of Physics and Astronomy, \\ Northwestern University, Evanston, IL 60208, USA\\
  $^{13}$Department of Physics, University of California, Davis, 1 Shields Ave, Davis, CA 95616, USA\\ 
  $^{14}$The Oskar Klein Centre, Department of Astronomy, AlbaNova, SE-106 91 Stockholm, Sweden\\  
  $^{15}$Instituto de Astrof{\'{\i}}sica and Centro de Astroingenier{\'{\i}}a, Facultad de F{\'{i}}sica, Pontificia Univ. Cat{\'{o}}lica de Chile, Casilla 306, Santiago 22, Chile \\
  $^{16}$Space Science Institute, 4750 Walnut Street, Suite 205, Boulder, Colorado 80301, USA\\
  $^{17}$INAF- Osservatorio Astronomico di Capodimonte, Salita Moiariello 16, 80131 Napoli, Italy\\  
  $^{18}$Max-Planck-Institut f{\"u}r Extraterrestrische Physik, Giessenbachstra\ss e 1, 85748, Garching, Germany\\
  $^{19}$Department of Particle Physics and Astrophysics, Weizmann Institute of Science, Rehovot 761000, Israel \\  
  $^{20}$Fundaci\'on Galileo Galilei, INAF Telescopio Nazionale Galileo, Rambla Jos\'e Ana Fern\'andez P\'erez 7, 38712 Bre\~na Baja, TF, Spain\\ 
  $^{21}$Las Cumbres Observatory, 6740 Cortona Dr Suite 102, Goleta, CA 93117-5575, USA\\
  $^{22}$Department of Physics, University of California, Santa Barbara, CA 93106-9530, USA\\  
  $^{23}$Applied Physics Department, Polytechnic Engineering School of Algeciras. \\ University of C\'adiz, Avenida Ram\'on Puyol s/n, 11202 Algeciras, Spain\\
      }
\date{Released 2017 Sept 15}

\pagerange{\pageref{firstpage}--\pageref{lastpage}} \pubyear{2017}

\begin{document}

\label{firstpage}

\maketitle

\begin{abstract}

We present one year of optical and near-infrared photometry and spectroscopy of the Type IIP SNe~2013K and 2013am. Both objects are affected by significant extinction, due to their location in dusty regions of their respective host galaxies, ESO 009-10 and NGC 3623 (M65). From the photospheric to nebular phases, these objects display spectra congruent with those of underluminous Type~IIP SNe (i.e. the archetypal SNe~1997D or 2005cs), showing low photospheric velocities ($\sim2 \times 10^{3}$ km s$^{-1}$ at 50~d) together with features arising from Ba~II which are particularly prominent in faint SNe IIP. The peak $V$-band magnitudes of  SN~2013K ($- 15.6$ mag) and SN~2013am ($- 16.2$ mag) are fainter than standard-luminosity Type~IIP SNe. The ejected Nickel masses are $0.012\pm0.010$ and $0.015\pm0.006$~M$_{\odot}$ for SN~2013K and SN~2013am, respectively. The physical properties of the progenitors at the time of explosion are derived through hydrodynamical modelling. Fitting the bolometric curves, the expansion velocity and the temperature evolution, we infer total ejected masses of 12 and 11.5~M$_{\odot}$, pre-SN radii of $\sim460$ and $\sim360$~R$_{\odot}$, and explosion energies of 0.34 foe and 0.40 foe for SN~2013K and SN~2013am. Late time spectra are used to estimate the progenitor masses from the strength of nebular emission lines, which turn out to be consistent with red supergiant progenitors of $\sim15$ M$_{\odot}$. For both SNe, a low-energy explosion of a moderate-mass red supergiant star is therefore the favoured scenario.
\end{abstract}


\section{Introduction}

Type~II plateau supernovae (SNe IIP) are thought to be the explosive end-stages of H-rich massive stars (above 8 M$_\odot$; see e.g. Woosley \& Weaver 1986; Heger et al. 2003; Pumo et al. 2009). From an observational point of view, these SNe are characterised by the presence of broad hydrogen lines with P-Cygni profiles in their spectra, and a long-lasting ($\sim100$~d) plateau in the light curve, while the hydrogen recombination wave propagates inside the SN ejecta. At the end of the plateau phase, there is a sudden drop of luminosity in the bolometric curve to meet the radioactive tail, which is characterised by a linear decline of 0.98 mag~(100~d)$^{-1}$, and where the electromagnetic emission is powered by the decay of $^{56}$Co to $^{56}$Fe. 

Numerical simulations of such photometric properties suggest that Type IIP SNe originate from red-supergiant stars \citep[RSGs; see for example][ and references therein]{grass:1971,falk:1977,woosley:1986,heger:2003,utrobin:2007,pumo:2011,pumo:2013,dessart:2013}. This association has been confirmed by the detection of several RSGs as precursors of Type IIP SNe in pre-explosion images (e.g. Smartt 2009 for a review). However, some disagreement on progenitor masses remains, and there is a general trend for masses coming from hydrodynamical modelling to be higher than those determined from pre-SN imaging \citep[see ][]{utrobin:2008,smartt:2009b,smartt:2015b}. 
Furthermore, it is not known how the observed diversity in both photometric and spectroscopic properties of SNe IIP depends on their progenitor properties \citep[][]{hamuy:2003,dessart:2013,anderson:2014,faran:2014,sanders:2015,galbany:2016,valenti:2016,rubin:2016,galyam:2017}. At early epochs, there is a wide range between standard-luminosity SNe IIP \citep[characterised by an average peak magnitude of $M_V = -16.74$ mag with $\sigma = 1.01$, see][]{anderson:2014} and low-luminosity events, down to $\sim -15$ mag \citep{pastorello:2004,spiro:2014}. SN~1997D was the first underluminous SN~IIP (Turatto et al. 1998, Benetti et al. 2001), followed by a growing number of similar events \citep{hamuy:2003,pastorello:2004,pastorello:2009,utrobin:2007,fraser:2011,vandyk:2012,arcavi:2013,spiro:2014}. 
In the ``middle ground'' only few intermediate-luminosity SNe have been studied 
\citep[i.e. SN~2008in, SN~2009N, SN~2009js, SN~2009ib, SN~2010id, SN~2012A, see][ respectively]{roy:2011,takats:2014,gandhi:2013,takats:2015,galyam:2011,tomasella:2013}. 

Typically, these intermediate-luminosity SNe also show low expansion velocities that match those of the extremely faint SNe~IIP, while their late time light curves indicate $^{56}$Ni masses ranging between those of the Ni-poor underluminous objects (less than $10^{-2}$ M$_{\odot}$ of $^{56}$Ni) and more canonical values ($0.06$ to $0.10$ M$_{\odot}$ of $^{56}$Ni; cf. M\"uller et al. 2017, Anderson et al. 2014).
Overall, the observational properties of intermediate SNe IIP suggest a continuous distribution of Type IIP properties. 
The cause of the observed spread of parameters among faint, intermediate and standard SNe IIP still remains unclear.

In this paper we present observational data and hydrodynamical modelling for a pair of intermediate-luminosity objects, SN~2013K and SN~2013am. These events first caught our interest due to their relatively low expansion velocities and intrinsic magnitudes. The earliest classification spectra of SN~2013K \citep{taddia:2013} and SN~2013am \citep{benetti:2013} were similar to the underluminous Type~IIP SN 2005cs \citep{pastorello:2006,pastorello:2009}, showing ejecta velocities of about 6300 and 8500 km s$^{-1}$, respectively, from the H${\beta}$ absorption minima. For both SNe, we immediately started extensive campaigns of photometric and spectroscopic monitoring, lasting over one year. These campaigns were enabled by the Public ESO Survey of Transient Objects (PESSTO, Smartt et al. 2015) and the Asiago Classification Program \citep[ACP,][]{tomasella:2014}, supported also by other facilities.
Over the course of the photometric follow-up campaign,
it became apparent that the luminosity and duration of the plateau-phase for SNe~2013K and 2013am are closers to those of the intermediate-luminosity Type IIP SNe, matching SN 2012A \citep{tomasella:2013} 
rather than more extreme sub-luminous Type IIP SNe. 
We note that optical data for SN~2013am (complemented by public {\it Swift} $UV$ photometry) have been previously reported by \cite{zhang:2014}.

Our motivation for a comparative study of the almost identical SNe~2013K and 2013am, is to expand the sample of well-studied intermediate-luminosity events in the literature, and with that to help understanding the physical causes behind the observed diversity in Type IIP SNe. To this end, we determine the physical parameters of the progenitors at the point of explosion through hydrodynamical modelling of the SN observables \citep[i.e. bolometric light curve, evolution of line velocities and continuum temperature at the photosphere, see][]{pumo:2010,pumo:2011,pumo:2017}. Also, we use the observed nebular spectra, and in particular the luminosities of the forbidden lines of Oxygen and Calcium or Nickel and Iron, to constrain the main-sequence mass of the progenitors \citep{fransson:1989,maguire:2012,jerkstrand:2012,jerkstrand:2015}.

The paper is organised as follows: in Section~\ref{2} we report basic information on the detection of SN~2013K and SN~2013am. Optical, near-infrared (NIR), ultraviolet ($UV$) observations together with a description of the data reduction process are provided in Section~3. In Section~4, we present the optical and NIR photometric evolution of the two SNe, comparing their colour and bolometric light curves with those of other Type IIP SNe, deriving the ejected Nickel masses from the bolometric radioactive tail, and analysing the spectroscopic data from photospheric to nebular phases. Section~5 is devoted to hydrodynamical modelling. Finally, in Section~6 we discuss and summarise the main results of our study.

\section{The two SNe and their host galaxies}\label{2}

In Table~\ref{info}, we summarise the main observational data for SNe~2013K and 2013am and their host galaxies. 
The Tully-Fisher distance moduli reported in the \cite{nasonova:2011} catalogue\footnote{retrieved from {\it The Extragalactic Distance Database} http://edd.ifa.hawaii.edu} of $32.66 \pm 0.40$ mag for SN 2013K and $30.54 \pm 0.40$ mag for SN 2013am (adopting $H_0$ = $73 \pm 5$ km s$^{-1}$ Mpc$^{-1}$) are used throughout this paper. The foreground Galactic extinctions $A_B$ = 0.516 mag and $A_B$ = 0.090 mag adopted for SNe~2013K and 2013am, respectively, are from Schlafly \& Finkbeiner (2011). For the estimation of the total extinction (Galactic plus host galaxy) see Section~\ref{extinction}.

\subsection{SN~2013K}

The discovery of SN~2013K, close to the nucleus of the southern galaxy ESO 009-10 (Fig.~1), was reported by S. Parker (Backyard Observatory Supernova Search - BOSS)\footnote{http://www.bosssupernova.com} on 2013 Jan. 20.413 UT (UT will be used hereafter in the paper). The transient was classified by \cite{taddia:2013} on
behalf of the PESSTO collaboration \citep[see][]{smartt:2015} as a Type~II SN a few days past maximum light. The explosion epoch is not well defined, as the closest non-detection image was 
taken by S.~Parker on 2012 Dec. 9.491 at a limiting magnitude 18. The template-matching approach applied to the early spectra, using the GELATO and SNID  spectral classification tools \citep[][]{avik:2008,blondin:2007}, and the light-curve, allow to constrain the explosion epoch to be around 2013 Jan. 9, with a moderate uncertainty (MJD\,=\,$56302.5^{+5}_{-5}$). After the classification, we promptly triggered a PESSTO follow-up campaign on this target.

\subsection{SN~2013am}

SN 2013am was first detected by \cite{nakano:2013}  in M65 (NGC~3623, see Fig.~2) on 2013 Mar. 21.638 UT. It was classified as a young SN~II by \cite{benetti:2013} under the Asiago Classification Program \citep[ACP,][]{tomasella:2014}.  
There was no evidence of the SN (down to an unfiltered magnitude $\sim$19) on frames taken by the Catalina Real-time Transient Survey (CRTS) on Mar. 20.198, indicating that the SN was caught very early. This stringent non-detection constrains the explosion time with a small uncertainty. In this paper, we adopt Mar. 21.0 (MJD\,=\,$56371.5^{+1.5}_{-1.0}$) as the explosion epoch. 
After the classification, we initiated a joint PESSTO and Asiago programme follow-up campaign on this target.

\begin{table*}
\caption{Main data for SNe 2013am and 2013K, and the respective host galaxies, M~65 and ESO 009-10.}\label{info}
\begin{tabular}{llll}
\hline \\
 &  SN 2013am & SN 2013K \\
 \hline \\
Host galaxy  & M~65 & ESO 009-10 \\
Galaxy type     & SABa & SAbc\\
Heliocentric velocity  (km s$^{-1}$) & $807\,\pm\,3$   & $2418\,\pm\,10$\\
Distance (Mpc) &12.8 & 34.0 \\
Distance modulus (mag) & $30.54\,\pm\,0.40$ & $32.66\,\pm\,0.40$\\
Galactic extinction $A_{B}$ (mag)   & 0.090 & 0.516\\
Total extinction $A_{B}, A_{V}, E(B-V)$ (mag) & $\approx$2.5, 2.0, $0.65\pm0.10$ & $\approx$1.0, 0.7, $0.25\pm0.20$ \\
                                   & &\\
 SN Type & IIP &IIP \\
 RA(J2000.0) & 11$^h$18$^m$56.95$^s$ & 17$^h$39$^m$31.54$^s$ \\                                   
 Dec(J2000.0) &+13$^{\circ}$03$'$49\farcs4 & $-85^{\circ}$18$'$38\farcs1 \\
 Offset from nucleus & 15\farcs E 102\farcs S & 6\farcs E 1\farcs S \\
 Date of discovery UT & 2013 Mar. 21.64 & 2013 Jan. 20.41\\
 Date of discovery (MJD)                  &  56372.6       &    56312.4 \\
 Estimated date of explosion (MJD) & $56371.5^{+1.5}_{-1.0}$ & $56302.0^{+5}_{-5}$\\
 $m_V$ at maximum (mag) & $16.34\,\pm\,0.01$ & $17.67\,\pm\,0.04$ \\ 
  $M_V$ at maximum (mag) &  $-16.2\,\pm\,0.3$& $-15.6\,\pm\,0.2$\\                                     
 $L_{\rm bol}$ peak ($\times 10^{41}$ erg s$^{-1}$) & 15.0$^{+2.2}_{-2.0}$  &  5.2$^{+2.5}_{-1.6}$ (\textit{UV} missing)\\                                   
\hline \\
\end{tabular}
\end{table*}

\begin{figure}
\includegraphics[scale=.47,angle=0]{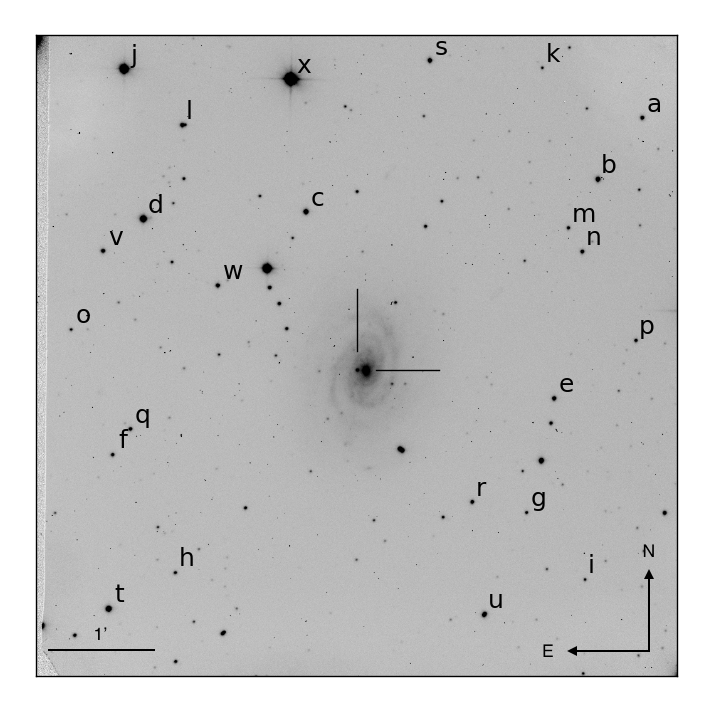}
\caption{SN 2013K (located 6 arcsec East and 1 arcsec South of the center of ESO 009-10) and local sequence stars (CTIO SMARTS 1.3m Telescope, $R$-band image obtained on 2013 Feb. 12, with exposure time 150s).} 
\label{map}
\end{figure}

\begin{figure}
\includegraphics[scale=.47,angle=0]{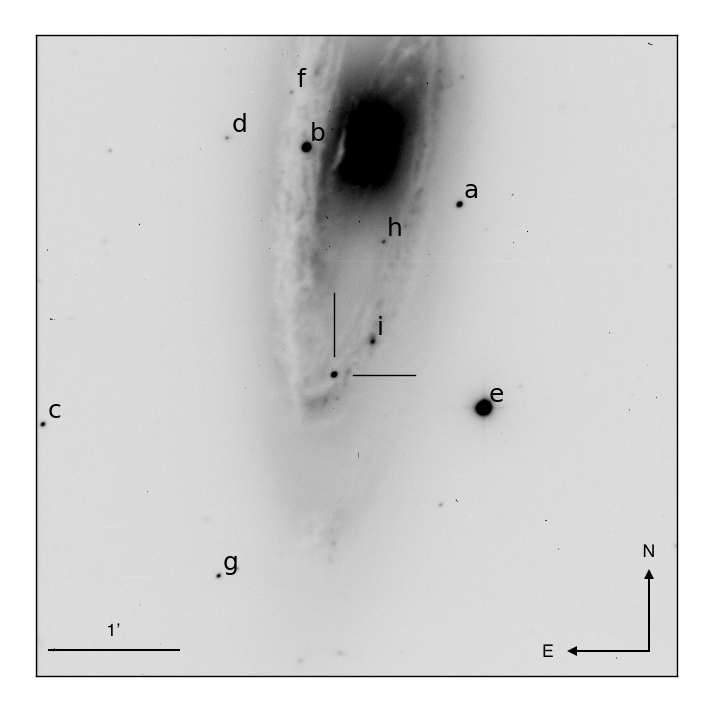}
\caption{SN 2013am and local sequence stars (Liverpool Telescope, $g$-band image obtained on 2013 Mar. 30, with exposure time 180s).} 
\label{map}
\end{figure}

\section{Observations and data reduction}

\subsection{Photometry}\label{photo}

\begin{table*}
\caption{List of observing facilities employed for optical and infrared photometry.} \label{telescopes_phot}
\begin{tabular}{lllcc}
\hline
Telescope                        & Instrument & Site & FoV  & Scale \\
                                         &                     &        & [arcmin$^{2}$] & [arcsec pix$^{-1}$]\\
\hline
\multicolumn{5}{c}{ \bf Optical facilities }\\
Schmidt 67/92cm         & SBIG          & Asiago, Mount Ekar (Italy) &  $57\times38$ &  0.86 \\
Copernico 1.82m        & AFOSC      & Asiago, Mount Ekar (Italy) &   $8 \times 8$   &  0.48 \\
Prompt   41cm              & PROMPT  & CTIO Observatory (Chile)  &    $11 \times 11$  & 0.59 \\
SMARTS 1.3m            & ANDICAM-CCD & CTIO Observatory (Chile)  &    $6 \times 6$  & 0.37 \\
LCO 1.0m               & kb73, kb74 &  CTIO Observatory (Chile)  &  $16 \times 16$  & 0.47 \\
LCO 1.0m               & kb77 &  McDonald  Observatory, Texas (USA) &    $ 16 \times 16$  & 0.47 \\
LCO FTS 2.0m       & fs01 &  Siding Spring  (Australia)&  $10 \times 10$  & 0.30 \\
LCO FTN 2.0m       & fs02 &  Haleakala, Hawai (USA) &  $10 \times 10$  & 0.30 \\
Liverpool 2.0m LT         & RATCam &  Roque de los Muchachos, La Palma, Canary Islands (Spain)    &  $4.6 \times 4.6$ & 0.135\\
Trappist 60cm            & TRAPPISTCAM & ESO La Silla Observatory (Chile) &  $ 27 \times 27$ & 0.65 \\
ESO NTT   3.6m           & EFOSC2            & ESO La Silla Observatory (Chile) & $ 4 \times 4$  &  0.24 \\
TNG    3.6m                  &  LRS &  Roque de los Muchachos, La Palma, Canary Islands (Spain) &$8 \times 8 $ & 0.25 \\
\multicolumn{5}{c}{ \bf Infrared facilities }\\
REM      60cm             & REMIR & ESO La Silla Observatory (Chile) &  $10 \times 10$ & 1.22 \\
ESO NTT   3.6m           & SOFI          & ESO La Silla Observatory (Chile) & $5 \times 5 $  &  0.29 \\
NOT     2.56m              & NOTCam     & Roque de los Muchachos, La Palma, Canary Islands (Spain) & $4 \times 4 $ & 0.234\\
SMARTS 1.3m            & ANDICAM-IR & CTIO Observatory (Chile)  &    $2.4 \times 2.4$  & 0.276 \\ \hline
\end{tabular}
\end{table*}

Optical and near infrared (NIR) photometric monitoring of SNe 2013am and 2013K was obtained using multiple observing facilities, summarised in Table~\ref{telescopes_phot}. For SN 2013am, we collected  data using Johnson-Cousins \textit{UBVRI}, Sloan \textit{ugriz} plus \textit{JHK} filters. For SN 2013K mostly \textit{BVRI}-band images were taken, with only three epochs in $U$ (obtained with NTT$+$EFOSC2; photometric standard fields were also observed during these nights), and four in \textit{gri} bands. The latter, covering the critical phase from the end of the plateau to the beginning of the radioactive tail, were transformed to \textit{VRI} (Vega) magnitudes using relations from \cite{chonis:2008}. 

All frames were pre-processed using standard procedures in {\sc iraf} for bias subtraction, flat fielding and astrometric calibration. For NIR, illumination correction and sky background subtraction were applied. For later epochs, multiple exposures obtained in the same night and with the same filter were combined to improve the signal-to-noise ratio. The photometric calibration of the two SNe was done relative to the sequences of stars in the field (Figs~1 and 2) and calibrated using observations either of Landolt (1992) or Sloan Digital Sky Survey (SDSS, Data Release 12, Alam et al. 2015)\footnote{http://www.sdss.org} fields. The local sequences (see Table~\ref{local}) were used to compute zero-points for non-photometric nights. In the NIR, stars from the 2MASS catalogue were used as photometric reference.
The SN magnitudes have been measured via point-spread-function (PSF) fitting using a dedicated pipeline ({\sc snoopy} package, Cappellaro 2014). {\sc snoopy} is a collection of {\sc python} scripts calling standard {\sc iraf} tasks (through {\sc pyraf}) and specific data analysis tools such as {\sc sextractor} for source extraction and {\sc daophot} for PSF fitting. The sky background at the SN location is first estimated with a low-order polynomial fit of the surrounding area. Then, the PSF model derived from isolated field stars is simultaneously fitted to the SN and any point source projected nearby (i.e. any star-like source within a radius of $\sim5\times$FWHM from the SN). The fitted sources are removed from the original images, an improved estimate of the local background is derived and the PSF fitting procedure iterated. The residuals are visually inspected to validate the fit.
Error estimates were derived through artificial star experiments. In this procedure, fake stars with magnitudes similar to the SN are placed in the fit residual image at a position close to, but not coincident with, the SN location. The simulated image is processed through the same PSF fitting procedure and the standard deviation of the recovered magnitudes of a number of artificial star experiments is taken as an estimate of the instrumental magnitude error. For a typical SN, this is mainly a measure of the uncertainty in the background fitting. The instrumental error is combined (in quadrature) with the PSF fit error, and the propagated errors from the photometric calibration chain.

Johnson-Bessell, Sloan optical magnitudes and NIR photometry of both SNe (and associated errors) are listed in Tables~\ref{phot1}, \ref{phot2}, \ref{phot3}, \ref{phot4}, \ref{phot5} and \ref{phot6}. Magnitudes are in the Vega system for the Johnson-Bessell filters, and in the AB system for the Sloan filters.

An alternative technique for transient photometry is template subtraction. However, this requires the use of exposures of the field obtained before the SN explosion or after the SN has faded below the detection threshold. The template images should be taken with the same filter, and with good signal to noise and seeing. In principle, they should be obtained with the same telescope and instrumental set-up, but in practice we are limited to what is actually available in the public archives. We retrieved pre-discovery SDSS \textit{ri}-band exposures covering M65, the host of SN~2013am. The template (SDSS) images were geometrically registered to the same pixel grid as the SN ones, and the PSFs matched by means of a convolution kernel determined from reference sources in the field. Then, the template image was subtracted from the SN frame and the difference image was used to measure the transient magnitude. Comparing the PSF fitting vs. template subtraction, the measured values differ by less than 0.1 mag. We considered this to be a satisfactory agreement given the differences between the passband of template (Sloan \textit{ri}) and SN images (Johnson-Bessell \textit{RI}). We conclude that our PSF fitting magnitudes are properly corrected for background contamination at least in the case of SN 2013am.

The light curves of SN~2013am were complemented with $UV$-optical photometry  (at eleven epochs, from phases $+2$ d to $+29$ d) obtained with the Ultra-Violet/Optical Telescope \citep[UVOT;][]{Roming:2005} onboard the {\it Swift} spacecraft \citep{Gehrels:2004}. The data were retrieved from the Swift Data Center\footnote{https://swift.gsfc.nasa.gov/sdc/}  
and were re-calibrated in 2016 using version 2015.1 of Peter Brown's photometry pipeline and version {\sc swift\_Rel4.5(Bld34)\_27Jul2015} of HEASOFT  \citep{brown:2014,brown:2015}. 
The reduction is based on the work of \cite{brown:2009}, including subtraction of the host galaxy count rates and uses the revised $UV$ zero-points and time-dependent sensitivity from \cite{breeveld:2011}. 
We note that the photometry of \cite{zhang:2014} for SN 2013am is systematically brighter in \textit{RI} during the radioactive tail phase. The reason for this discrepancy is unclear, and we only include the photometry earlier than 109~d from \citeauthor{zhang:2014} when computing the pseudo-bolometric light curve.

\subsection{Spectroscopy}

The journals of spectroscopic observations for SNe~2013K and 2013am, both optical and NIR, are reported in Tables~\ref{telescope_spec} and \ref{telescope_spec2} respectively.

Data reduction was performed using standard {\sc iraf} tasks. First, images were bias and flat-field corrected. Then, the SN spectrum was extracted, subtracting the sky background along the slit direction.
One-dimensional spectra were extracted weighting the signal by the variance based on the data values and a Poisson/CCD model using the gain and readout-noise parameters.
The extracted spectra have been wavelength-calibrated using comparison lamp spectra and flux-calibrated using spectrophotometric standard stars observed, when possible, in the same night and with the same instrumental configuration as the SN. The flux calibration of all spectra was verified against photometric measures and, if necessary, corrected. The telluric absorptions were corrected using the spectra of both telluric and spectrophotometric standards. 

All PESSTO spectra collected with the 3.6m New Technology Telescope (NTT+EFOSC2 or NTT+SOFI, cf. Tables~\ref{telescope_spec} and \ref{telescope_spec2}) are available through the ESO Science Archive Facility. Full details of the formats of these spectra can be found on the PESSTO website\footnote{www.pessto.og} and in \cite{smartt:2015}. Some spectra were obtained under the ANU WiFeS SuperNovA Programme, which is an ongoing supernova spectroscopy campaign utilising the Wide Field Spectrograph on the Australian National University 2.3-m telescope. The first and primary data release of this programme \citep[AWSNAP-DR1, see][]{childress:2016} releases 357 spectra of 175 unique objects collected over 82 equivalent full nights of observing from 2012 Jul. to 2015 Aug. These spectra have been made publicly available via the Weizmann Interactive Supernova data REPository (WISeREP, Yaron \& Gal-Yam 2012)\footnote{https://wiserep.weizmann.ac.il}.

\section{Data analysis}

\subsection{Light curves}\label{photo_evol}

\begin{figure}
\includegraphics[scale=.56,angle=0]{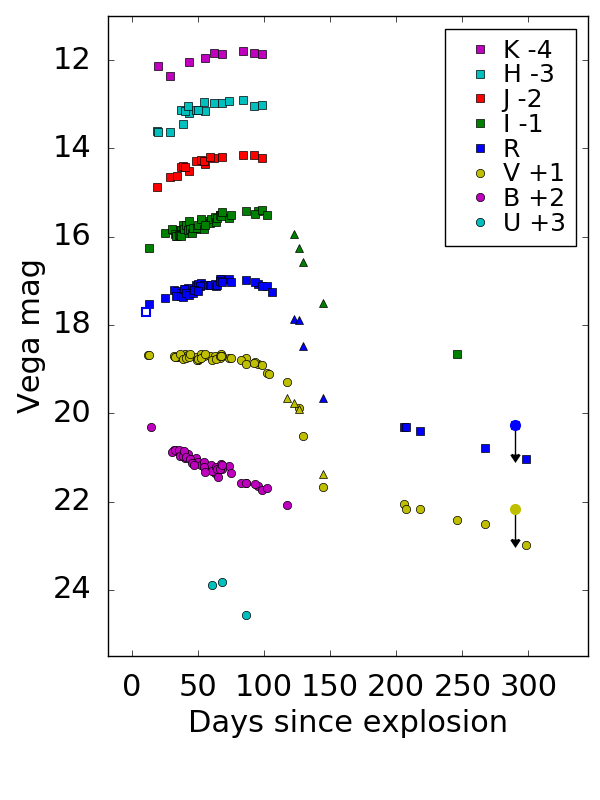}
\caption{Light curves of SN 2013K in the \textit{UBVRIJHK} bands. The triangles are \textit{gri}-band data converted to the Johnson-Cousins system (\textit{VRI}) using transformation formulas from \protect\cite{chonis:2008}. The empty square point on the $R$-band light curve is the unfiltered discovery magnitude from \protect\cite{taddia:2013}.} 
\label{lightcurve3}
\end{figure}

\begin{figure}
\includegraphics[scale=.56,angle=0]{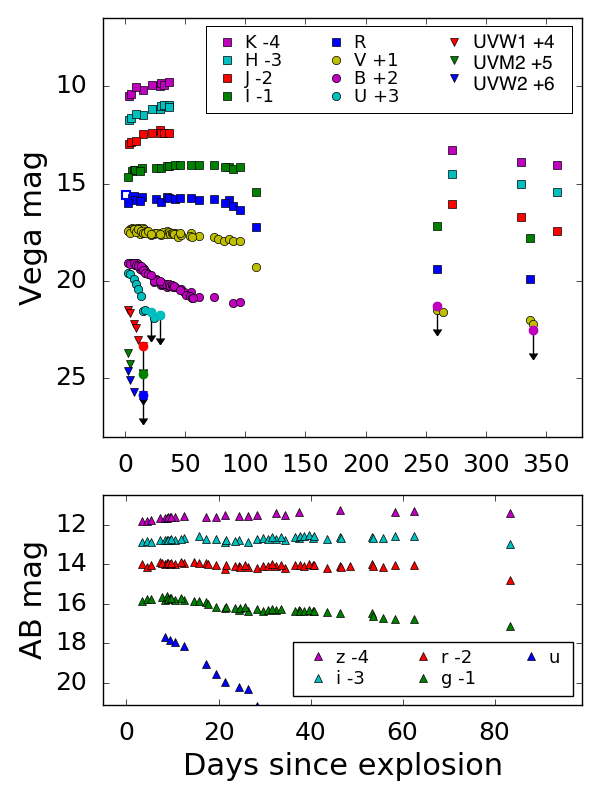}
\caption{Multiband \textit{UV} and optical light curve of SN 2013am. The {\it Swift} \textit{UVW1, UVM2, UVW2} bands are also included. The empty square point on the $R$-band light curve is the unfiltered discovery magnitude from \protect\cite{nakano:2013}.} 
\label{lightcurve1}
\end{figure}

The multicolour light curves of SN~2013K and SN~2013am are shown in Figs~\ref{lightcurve3} and \ref{lightcurve1}, respectively. The unfiltered discovery magnitudes are also included \citep{taddia:2013,nakano:2013} and plotted as empty squares on the $R$-band light curves. 
For SN~2013am, a polynomial fit to the early light curves shows that the $B$- and $V$-band maxima are reached about 5 and 8 days after the explosion at $m_B = 17.05 \pm 0.03$ mag and $m_V = 16.34 \pm 0.01$ mag, respectively, in fair agreement with the estimate by \cite{zhang:2014}. At subsequent epochs, we note a steep decline in the $B$ band ($\approx 3.76 \pm 0.13$ mag (100 d)$^{-1}$) during the first 50 d of evolution, and a moderate decline in the $V$ band ($\approx 0.68 \pm 0.05$ mag (100 d)$^{-1}$). The $R$ light curve shows a flat evolution during the first $\sim$ 85 d, with $m_R = 15.72 \pm 0.03$ mag, followed by the sharp luminosity drop from the plateau to the nebular phase. The $I$ band seems to flatten around day +26. Soon after, there is a re-brightening and a peak is reached at +58 d. A similar evolution (in \textit{VRI} bands) was noted by \cite{bose:2013} for SN~2012aw.

SN~2013K was discovered and classified about two weeks after the explosion, when the $B$-band light curve was already declining at a rate of $\approx$ 1.0 mag (100 d)$^{-1}$. 
The $V$-band light curve had already settled onto the plateau phase at the beginning of our follow-up campaign, during which the SN luminosity remained fairly constant at $m_V = 17.67 \pm 0.04$ mag for about 80 days. Between phases 20 and 70 d, both the $R$- and $I$-bands increased by about 0.8 mag (100 d)$^{-1}$, reaching a maximum shortly before the drop of luminosity which signs the end of the plateau. 
The duration of the plateau phase is around 95 days for SN~2013am and a dozen days longer for SN~2013K.  
For both these events, the rapid decline from the plateau ends at about 120 days, after a drop of $\sim 2$ mag in the $V$ band. A similar decline  
of two magnitudes was observed also for SN~2012A \citep{tomasella:2013} and SN~1999em \citep{elmhamdi:2003}, while underluminous objects can show deeper drop by about 3-5 mag \citep[][ see also Valenti et al. 2016 for a high-quality collection of Type II SN light-curves]{spiro:2014}.
Subsequently, the light curves enter the radioactive tail phase, during which there was a linear decline powered by the radioactive decay of $^{56}$Co into $^{56}$Fe. The decline rates in the \textit{VR} bands during the radioactive tail phase 
were $\gamma_{V} = 0.89 \pm 0.10$, $\gamma_{R} = 0.79 \pm 0.05$ mag (100 d)$^{-1}$ for SN~2013K $vs.$ $\gamma_{V} = 0.83 \pm 0.11$, $\gamma_{R}  = 0.86 \pm 0.09$ mag (100 d)$^{-1}$ for SN~2013am.  
For both SNe the slope of the pseudo-bolometric luminosity decline is close to the expected input from the $^{56}$Co  decay (0.98 mag 100 d$^{-1}$, see Section~\ref{bolometric} and Fig.~\ref{bol}). The decline rates obtained by \cite{zhang:2014} for SN~2013am are significantly flatter than our ones. Based on this data, they evoked a possible transitional phase in the light-curve evolution of SNe~IIP, where a residual contribution from recombination energy to the light curves prevents a steep drop on the radioactive tail, as already suggested by \cite{pastorello:2004,pastorello:2009} for underluminous SNe~1999eu and 2005cs \citep[see also ][]{utrobin:2007}. However, our late photometric measurements (obtained using both PSF fitting and template subtraction techniques, cf. Section~3.1) are fainter and decline faster than in \cite{zhang:2014}, and therefore we cannot confirm their finding.

\subsection{Extinction}\label{extinction}

In order to determine the intrinsic properties of the SN, a reliable estimate of the total reddening along the line of sight is needed, including the contribution of both the Milky Way and the host galaxy. 
The values of the foreground Galactic extinction derived from the \cite{schlafly:2011} recalibration of the \cite{schlegel:1998} infrared-based dust map are $E$($B-V$)$_{MW}$ = 0.022 mag for SN~2013am and $E$($B-V$)$_{MW}$ = 0.126 mag for SN~2013K. However, a relatively high contribution from the host galaxy is needed to explain the red colour of the spectral continuum,  especially for SN~2013am \citep{benetti:2013}.

To estimate the total reddening, we use the relation between the extinction and the equivalent width (EW) of the interstellar Na~I~D doublet \citep[e.g.][]{turatto:2003,poznanski:2012}, though we acknowledge that there is a large associated uncertainty due to the intrinsic scatter in this relation. We found that the Na~I~D absorption features can be detected in both SNe, with resolved host galaxy and Milky Way components. From the medium-resolution (2 \AA) spectra of SN~2013am obtained with the ANU 2.3m telescope (+WiFes spectrograph, phase +4.3 d and +7.6 d), we measured the EW($D_1 + D_2$) for the Galactic and host Na~I~D absorptions to be $\sim$0.24 \AA\ and $\sim$1.40 \AA\/, respectively. Applying the \cite{poznanski:2012} empirical relations (their eq.~9), we obtain $E$($B-V$)$_{MW}$ = 0.026$_{-0.004}^{+0.006}$ mag, in excellent agreement to the extinction from \cite{schlafly:2011} mentioned above, and $E$($B-V$)$_{host}$ = 0.63$_{-0.09}^{+0.13}$ mag. Applying a similar analysis to a lower resolution spectrum of SN~2013K, we find $E$($B-V$)$_{MW}$ = 0.10$_{-0.02}^{+0.04}$ mag (also in agreement with the value inferred from the infrared dust map), and a similar extinction inside the host galaxy. 

Overall, from the analysis of the Na~I~D lines we obtain moderate total extinction values, i.e. $E$($B-V$)$_{tot} \approx 0.65 \pm 0.10$  mag for SN~2013am ($A_V \approx 2$ mag, using the reddening law of Cardelli et al. 1989), and $E$($B-V$)$_{tot} \approx 0.25 \pm 0.20$  mag for SN~2013K ($A_V \approx 0.7$ mag), consistently with the location of the SNe in dusty regions of their host galaxies.
These values are adopted in the following analysis. 

We find that after applying these reddening corrections to the spectra, the GELATO spectral classification tool \citep{avik:2008} gives excellent matches to the low-velocity Type IIP SN~2005cs. We also note that our reddening estimate for SN~2013am is consistent, within the error, with the value derived by Zhang et al. (2014; cf. their Section 3.4).   

A consistency check of the adopted extinction values is based on the \cite{nugent:2006} correlation between the absolute magnitude of Type II SNe in $I$ band and the expansion velocity derived from the minimum of the Fe~II $\lambda$ 5169 P-Cygni feature observed during the plateau phase, at $t \approx +50$ d (their eq. 1). In Fig.~\ref{nugent} we plot the sample of nearby Type II SNe and the derived relation between absolute magnitude and photospheric velocity from Nugent et al. (2006; cf. their table~4).
In this figure, we add the data for five additional events: SNe~2005cs \citep{pastorello:2009}, 2009ib \citep{takats:2015}, 2012A \citep{tomasella:2013}, 2013K and 2013am (this work). After applying our adopted extinction correction, both SN 2013K and 2013am follow the expected relation.

\begin{figure}
\includegraphics[scale=.52,angle=0]{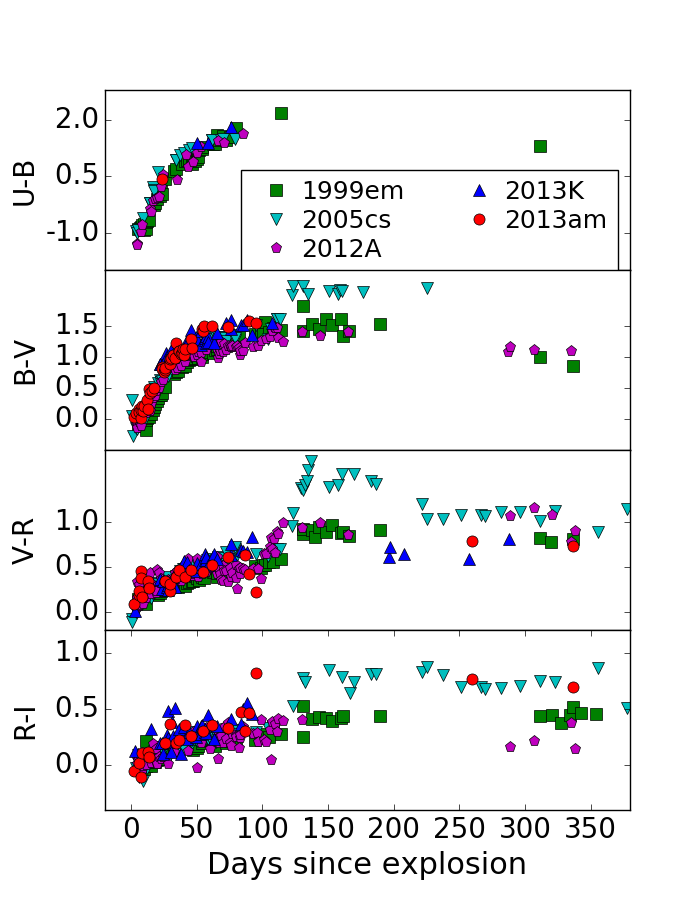}
\caption{From top to the bottom: $U - B$, $B - V$, $V - R$, $R - I$ colours of SNe 2013am and 2013K from early times to the nebular phase, compared to SNe 2005cs, 2012A and 1999em (see text for a discussion).} 
\label{col}
\end{figure}

\begin{figure}
\includegraphics[scale=.52,angle=0]{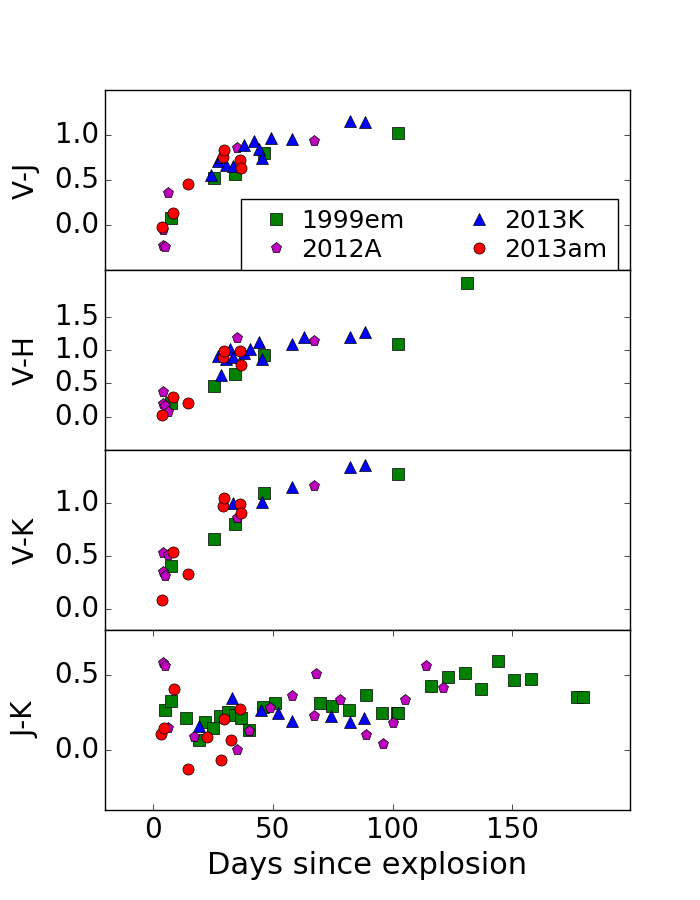}
\caption{NIR colour evolution of SNe 2013am and 2013K, compared to that of SNe 2012A and 1999em.} 
\label{col2}
\end{figure}

\begin{figure}
\includegraphics[scale=.37,angle=0]{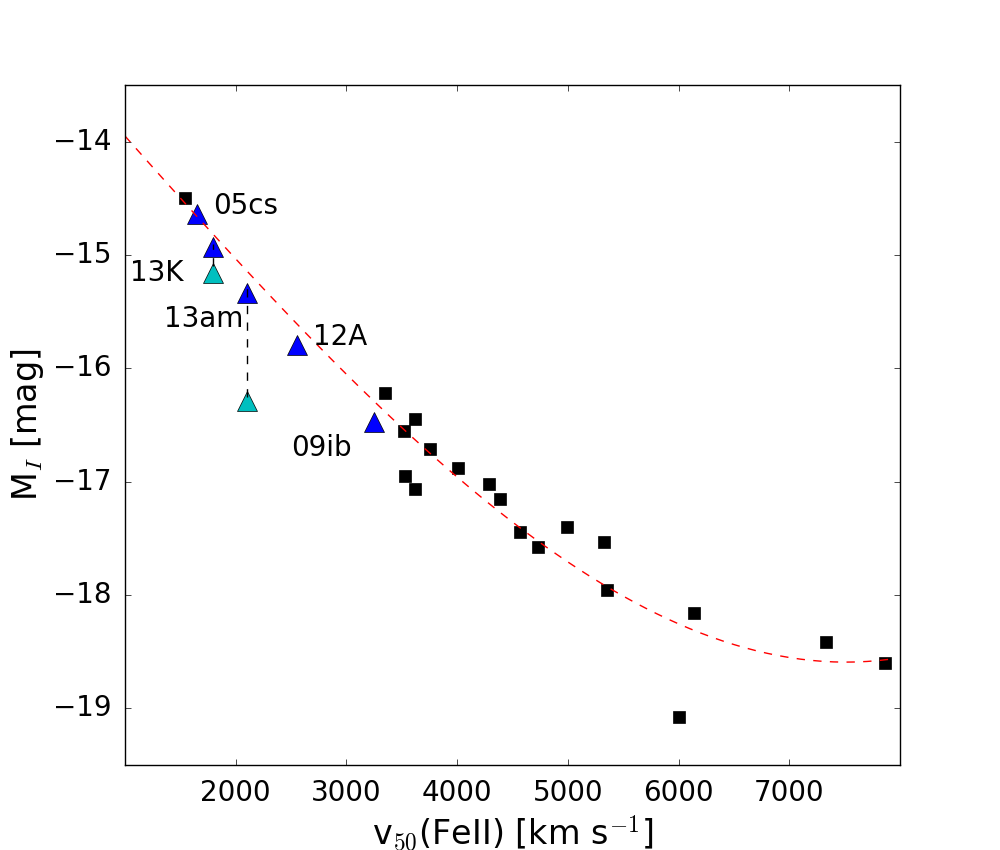}
\caption{Velocity-luminosity relation for a sample of nearby Type II SNe collected by Nugent et al. 2006 (black squares; their table~4). The fit to the data (dashed red line) is their eq. (1). The blue triangles are SNe~2005cs, 2009ib, and 2012A (Pastorello et al. 2009; Tak{\'a}ts et al. 2015; Tomasella et al. 2013), and SNe~2013K and 2013am (this work). For the latter two SNe, the cyan triangles are before applying the extinction correction.} 
\label{nugent}
\end{figure}

\subsection{Absolute magnitudes}\label{absmag}

With the above distances (Section~\ref{2}, Table~1), apparent magnitudes (Section~\ref{photo_evol}), and extinctions (Section~\ref{extinction}), we derive the following peak absolute magnitudes: 
$M_V \sim -15.6 \pm 0.2$, $M_R = -16.2 \pm 0.2$ for SN~2013K; $M_B = -16.1 \pm 0.2$, $M_V =  -16.2 \pm 0.3$ and $M_R = -16.4 \pm 0.3$ for SN~2013am.
The absolute magnitudes of both SNe are intermediate between the faint SN~2005cs \citep[$M_V = -15.1 \pm 0.3$ mag, cf.][]{pastorello:2006,pastorello:2009}, and normal Type IIP SNe, 
that have an average peak magnitude of $M_V = -16.74$ mag \citep[$\sigma = 1.01$; cf.][ see also Galbany et al. 2016]{anderson:2014}. The brightest normal Type IIP SNe can reach magnitudes around $\sim -18$ mag \citep{li:2011,anderson:2014}.

\subsection{Colour curves}

The optical and NIR colour curves of both SNe~2013K and 2013am after correction for the Galactic and host galaxy reddening (cf. Section~\ref{extinction}) are shown in Figs~\ref{col} and \ref{col2}. For a comparison, we also plot the colour evolution of the normal Type IIP SNe~1999em \citep{elmhamdi:2003}, 2012A \citep{tomasella:2013}, and the faint 2005cs \citep{pastorello:2009}. The common rapid colour evolution, especially in $B-V$, during the first month of evolution is due to the expansion and cooling of the photosphere. Between 100 and 150 days, both SN~2013am and SN~2013K show very little variation, similar to that experienced by the normal Type II SNe~2012A and 1999em, while the faint Type II SNe \citep[e.g. SN~2005cs and SN~2009md, see][ and references therein]{pastorello:2009,spiro:2014,fraser:2011} sometimes show a red peak in colour during the drop from the plateau phase. Contrary to the claim of Zhang et al. (2014; cf. their fig.~5) of an unusual $V-I$ red colour for SN~2013am during the nebular phase, the colour evolution for both SNe~2013am and 2013K is consistent with the normal Type II SNe, with no evidence of the red spike characterising the underluminous SNe~2005cs and 2009md five months after explosion.

\subsection{Pseudo-bolometric light curves and ejected Nickel masses}\label{bolometric}

The pseudo-bolometric luminosities of SNe~2013K and 2013am are obtained by integrating the available photometric data from the optical to the NIR. We adopt the following procedure: for all epochs we derived the flux at the effective wavelength in each filter. When observations for a given filter/epoch were not available, the missing values were obtained through interpolations of the light curve or, if necessary, by extrapolation, assuming a constant colour from the closest available epoch. The fluxes, corrected for extinction, provide the spectral energy distribution at each epoch, which is integrated by the trapezoidal rule, assuming zero flux at the integration boundaries. The observed flux is then converted into luminosity, given the adopted distance to each SN. The error, estimated by error propagation, is dominated by the uncertainties on extinction and distance.
The \textit{UV}-optical photometry retrieved from the {\it Swift} Data Center  
is also included when computing the pseudo-bolometric light curve of SN~2013am. Instead, \textit{UV} measurements are not available for SN~2013K and hence for this object the bolometric luminosity does not include this contribution.
At very early phases the far \textit{UV} emission contributes almost 50 per cent of the total bolometric luminosity, dropping to less than 10 per cent in about two weeks. This has to be taken into account when comparing SN~2013am with SN~2013K. Also, the striking diversity in the first $\sim$20 days of the light curves of Type IIP SNe may be attributed to the presence of a dense circumstellar material (CSM), as recently outlined in \cite{morozova:2017}. 
Colour corrections were applied to convert \textit{ubv} {\it Swift} magnitudes to the standard Landolt \textit{UBV} system.\footnote{http://heasarc.gsfc.nasa.gov/docs/heasarc/caldb/swift/docs/ uvot/uvot\_caldb\_coltrans\_02b.pdf}

\begin{figure}
\includegraphics[scale=.45,angle=0]{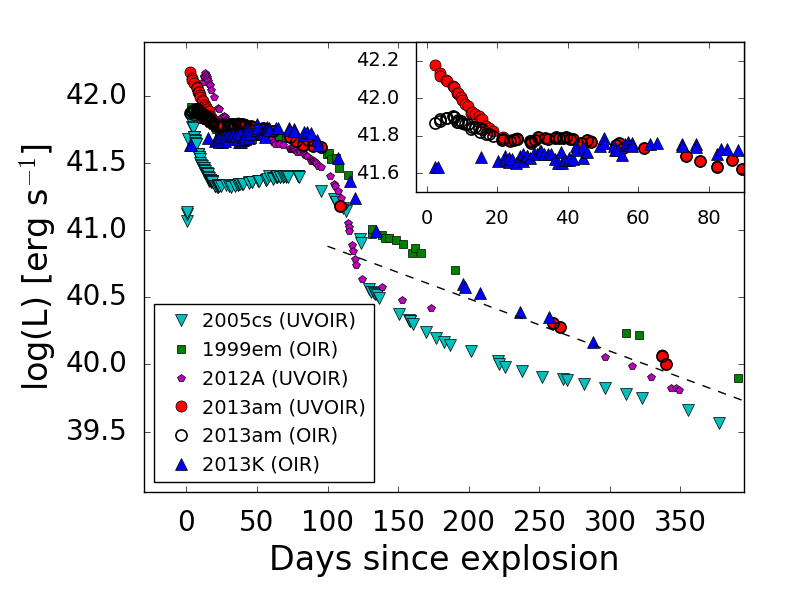}
\caption{Comparison of the \textit{OIR} or/and \textit{UVOIR} pseudo-bolometric light curves of SNe 2013K and 2013am with the low-luminosity SN 2005cs and the normal SNe 1999em and 2012A (see the figure legend and Section 4.5 for details). Some of the discrepancies between the curves of normal SNe 2013K, 2013am and 1999em at early times can be attributed to missing \textit{UV} flux in the pseudo-bolometric curves. The dashed line indicates the slope of the $^{56}$Co decay. } 
\label{bol}
\end{figure}

The pseudo-bolometric \textit{OIR} and \textit{UVOIR}\footnote{The abbreviation \textit{UVOIR} is used with different meanings in the literature. In this paper we use it to mean the flux integrated from 1600 \AA (Swift/UVOT \textit{UVW2}-band) to 25 \micron ($K$ band). If the integration starts from 3000 \AA (ground-based $U/u$-band) we use the label \textit{OIR}.}
light curves for SNe~2013K and 2013am respectively, are presented in Fig.~\ref{bol}, along with those of SNe 1999em (\textit{OIR}; adopting $\mu = 29.82 \pm 0.15$, $A_{B}^{tot} = 0.4 $ mag, Elmhamdi et al. 2003), 2005cs (\textit{UVOIR};  $\mu = 29.26 \pm 0.33$, $A_{B}^{tot} = 0.2$ mag, Pastorello et al. 2009) and 2012A (\textit{UVOIR}; $\mu = 29.96 \pm 0.15$, $A_{B}^{tot} = 0.15$ mag, Tomasella et al. 2013), which were computed with the same technique, including the {\it Swift}/UVOT contribution for SNe~2012A {\bf and 2005cs}, and using $H_0$ = 73 km s$^{-1}$ Mpc$^{-1}$. In Fig.~\ref{bol} we also plot the \textit{OIR} light curve for SN~2013am. The ratio between the \textit{UVOIR} and \textit{OIR} pseudo-bolometric luminosities of SN~2013am is around  2 at maximum light and decreases to 1.3 ten days after maximum. This is in agreement with the result obtained by \cite{bersten:2009} and \cite{faran:2018} for SN~1999em. Therefore we can assume that a similar correction should be applied to early phase of SN~2013K.

The early luminosity of SN~2013am matches that of SN~2012A, showing a peak luminosity of $L = 1.5^{+0.2}_{-0.2} \times 10^{42}$ erg s$^{-1}$ at about 2 days after explosion. We note a monotonic, rapid decline from the beginning of the follow-up campaign until 20 days from explosion. Later on, an almost constant plateau luminosity is reached, lasting about 20 days. In the successive 50 days, the light curve shows a slow monotonic decline, followed by a sudden drop, which marks the end of the hydrogen envelope recombination. The latter phase was not monitored as SN~2013am had disappeared behind the Sun. We recovered the SN at very late phases (starting from +261 d), in the linear tail phase, with decline rate close to that expected from the $^{56}$Co decay. 
The pseudo-bolometric peak of SN~2013K is reached close to the $R$-band maximum, and shortly before the drop from the plateau, with a luminosity of $L = 5.2^{+2.5}_{-1.6} \times 10^{41}$ erg~s$^{-1}$, however, the missing contribution in the \textit{UV} can represents a significant fraction of the total flux (up to a factor two), in the early days. The plateau duration and the subsequent drop to the radioactive tail are similar in both SN~2013K and SN~2013am. 

The luminosities during the radioactive tails of SNe~2013K and 2013am are comparable to those of SNe~2012A and 1999em, and significantly higher than that of the faint Type IIP SN~2005cs. Comparing the bolometric luminosities of SNe~2013K and 2013am with SN~1987A (from $+280$ to $+340$ d), we obtain a best fit for $L$(13K)$/L$(87A) $= 0.18\pm0.08$ and $L$(13am)$/L$(87A) = $0.20\pm0.03$. Thus, taking as reference the estimate of the  $^{56}$Ni mass for SN~1987A 
\citep[$0.075\pm0.005$ M$_\odot$,][]{danziger:1988,woosley:1989}, and propagating the errors, we obtain ejected  $^{56}$Ni  masses of $0.012\pm0.010$ and $0.015\pm0.006$ M$_{\odot}$ for SN~2013K and SN~2013am, respectively. Both these values are lower than the typical amount of $^{56}$Ni ejected by SNe IIP ($\sim 0.06-0.12$ M$_{\odot}$, Sollerman 2002, M\"uller et al. 2017; see also the distribution of $^{56}$Ni masses of Type II SNe sample by Anderson et al. 2014, showing a mean value of 0.033 M$\odot$ with 
$\sigma = 0.024$). However, these ejected  $^{56}$Ni masses are higher than those associated with Ni-poor ($\lesssim 10^{-2}$ M$_{\odot}$), low-energy events such as SNe~1997D \citep{turatto:1998,benetti:2001}, 2005cs \citep{pastorello:2006,pastorello:2009} and other faint SNe IIP \citep[cf. ][]{pastorello:2004,spiro:2014}.

\subsection{Optical spectra: from photospheric to nebular phases}\label{spec_evol}

\begin{figure*}
\includegraphics[scale=0.7,angle=0]{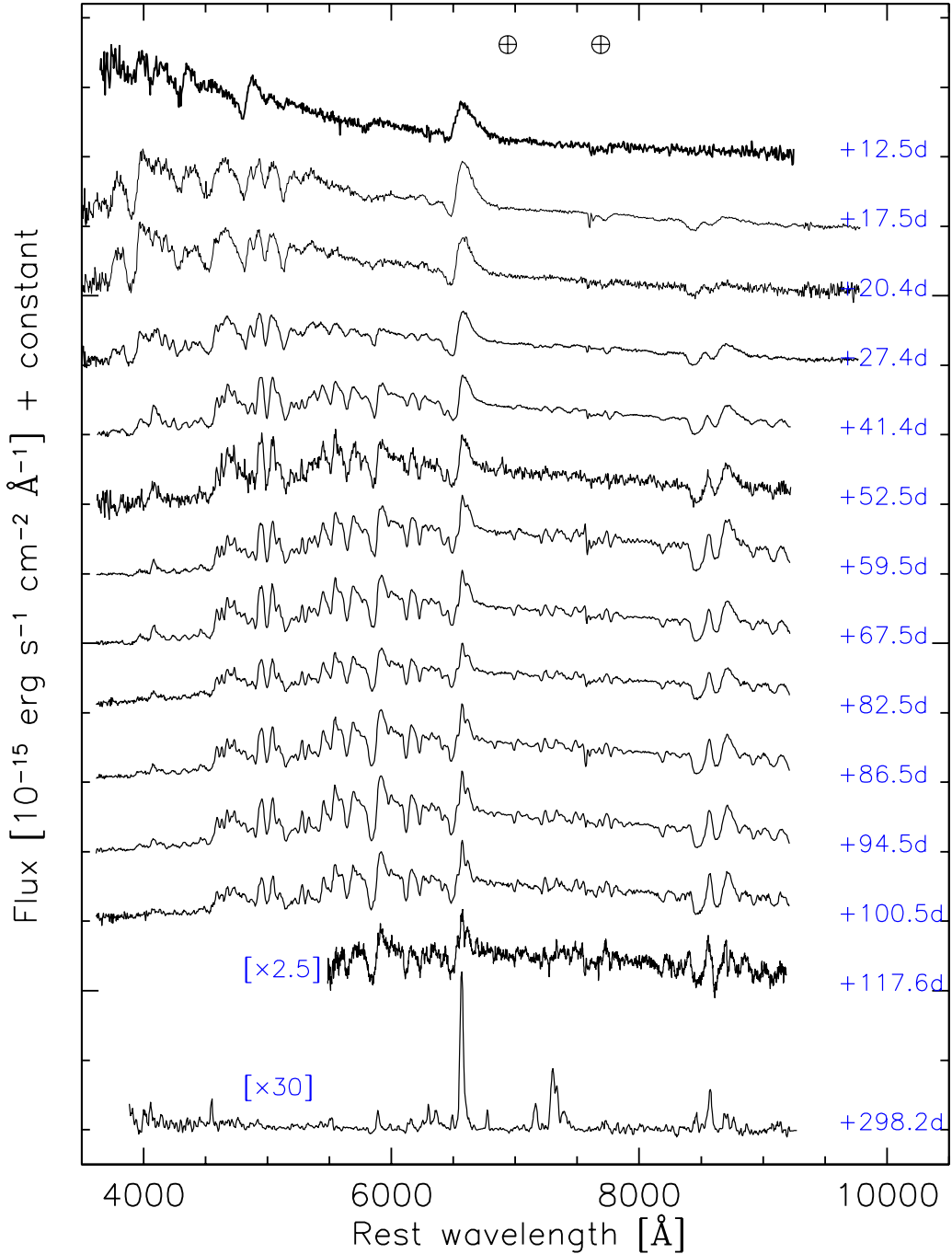}
\caption{SN 2013K: optical spectral evolution from the photospheric phase (+12.5~d),  to the nebular phase (+298.2~d). The spectra have been corrected for reddening and redshift, and shifted vertically for display. Phases (reported on the right of each spectrum) are relative to the explosion date, MJD = 56302.0$^{+5}_{-5}$. The positions of major telluric features (O$_{2}$ A $\&$ B) are marked with $\oplus$ symbols.} 
\label{fig_spectra2}
\end{figure*}

\begin{figure*}
\includegraphics[scale=0.7]{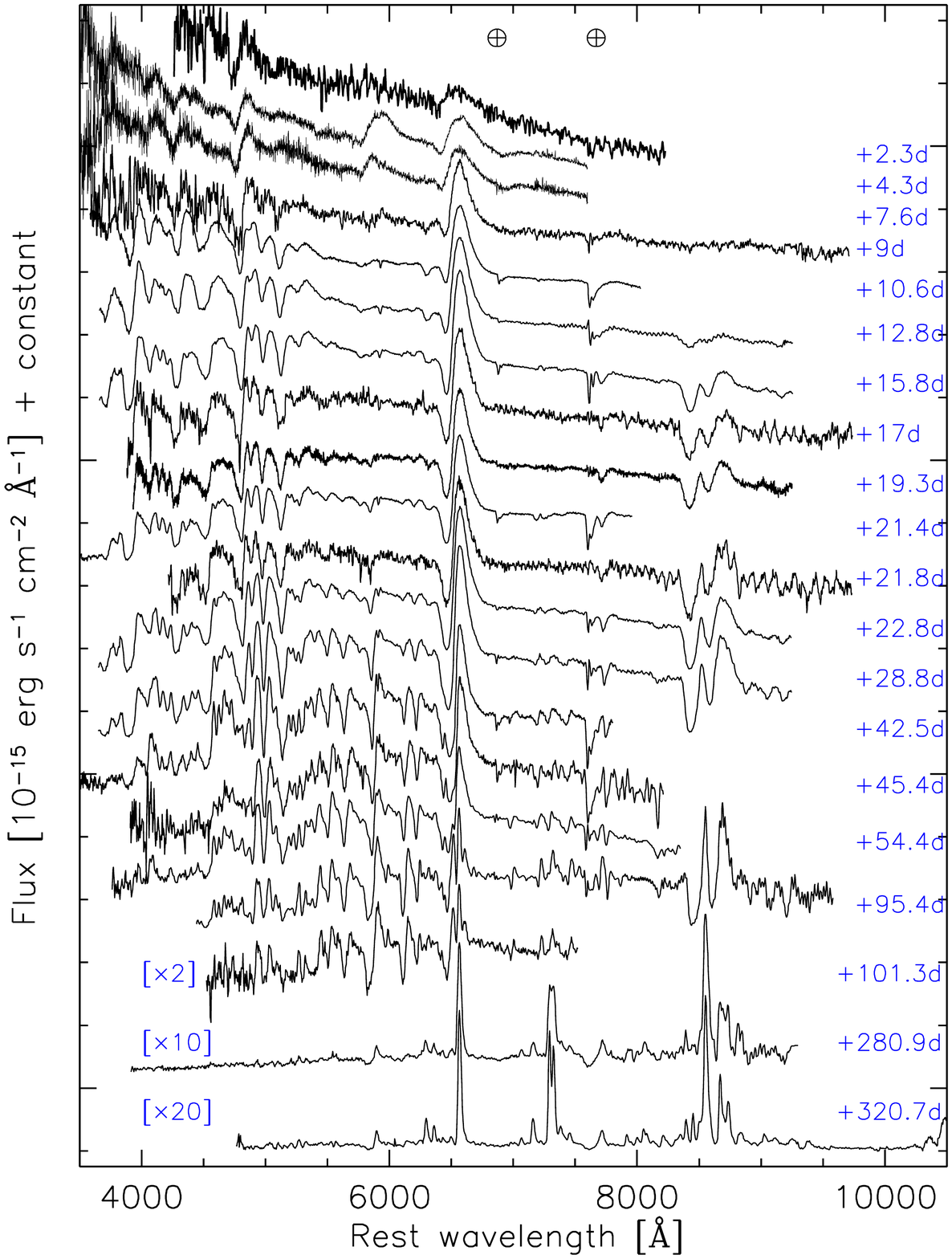}
\caption{SN 2013am: optical spectral evolution from the very early photospheric (+2.3 d),  to the nebular phase (+280.9~d and +320.7~d spectra). Only high signal-to-noise spectra are plotted here. The spectra have been corrected for reddening and redshift, and shifted vertically for clarity. Phases (reported on the right of each spectrum) are relative to the explosion date, MJD = 56371.5$^{+1.5}_{-1.0}$. Some spectra obtained at nearly the same epoch have been averaged in order to increase the signal-to-noise (e.g. the spectrum at phase +9~d is the average of three spectra taken at +7.8~d, +8.8~d, +9.8~d; the spectrum at phase +17~d is the average of spectra taken in two consecutive nights, +16.8~d and +17.8~d). The positions of major telluric features (O$_{2}$ A $\&$ B) are marked with $\oplus$ symbols.} 
\label{fig_spectra1}
\end{figure*}

In Figs~\ref{fig_spectra2} and \ref{fig_spectra1}, we present the entire spectral evolution of SNe~2013K and 2013am, from the photospheric to the nebular phases. For SN~2013am, the spectroscopic follow-up started shortly after the shock breakout. A total of 26 optical spectra  were taken, covering phases from $+$2.3~d to $+$320.7~d after the explosion. SN~2013K was caught about 12 days after explosion, and twelve spectra were collected, up until phase +298.2~d. The earlier spectra of both SNe are characterised by a blue continuum and prominent hydrogen Balmer lines with broad P-Cygni profiles. 
 
\subsubsection{Key spectral features}
  
Besides the hydrogen Balmer series, the first spectrum of SN~2013am (+2.3~d) shows a broad-line feature just blueward of H${\beta}$ which is tentatively associated with He~II $\lambda$ 4696 \citep[the presence of this line in Type IIP SNe~1999gi and 2006bp is extensively discussed by ][]{dessart:2008}. Two days later, He~I~$\lambda$ 5876 emerges. At this phase (+4.3~d), it is comparable in strength to H${\alpha}$, before weakening (+7.6 d) and disappearing soon after (+9 d). 
The first spectrum of SN~2013K, taken at phase +12.5 d, displays only a hint of He~I~$\lambda$ 5876. 

A few weeks later, the dominant features in both SNe~2013K and 2013am are metal lines with well developed, narrow P-Cygni profiles, arising from Fe~II ($\lambda$~4500, $\lambda\lambda\lambda$~4924, 5018, 5169, multiplet 42; these lines are also present in earlier spectra), Sc~II ($\lambda$ 4670, $\lambda$ 5031), Ba~II ($\lambda$~4554, $\lambda$~6142), Ca~II ($\lambda\lambda\lambda$~8498, 8542, 8662, multiplet 2, and H\&K), Ti~II (in a blend with Ca~II~H\&K), and Na~I~D (close to the position of the He~I~$\lambda$~5876). The permitted lines of singly ionised atoms of Fe and Ba begin to appear at phase $+$10.6 d for SN~2013am, and slightly later ($+$17 d) for SN~2013K. As the SNe evolve, their luminosity decreases, and their continua become progressively redder and dominated by metallic lines, such as Na~I~D and Ca~II~IR triplet. These features are clearly visible about one month after the explosion. In both SNe, strong line blanketing, especially due to Fe~II transitions, suppresses most of the \textit{UV} flux below 4000~\AA\/.

\subsubsection{Photospheric temperature}

\begin{figure}
\includegraphics[scale=.45,angle=0]{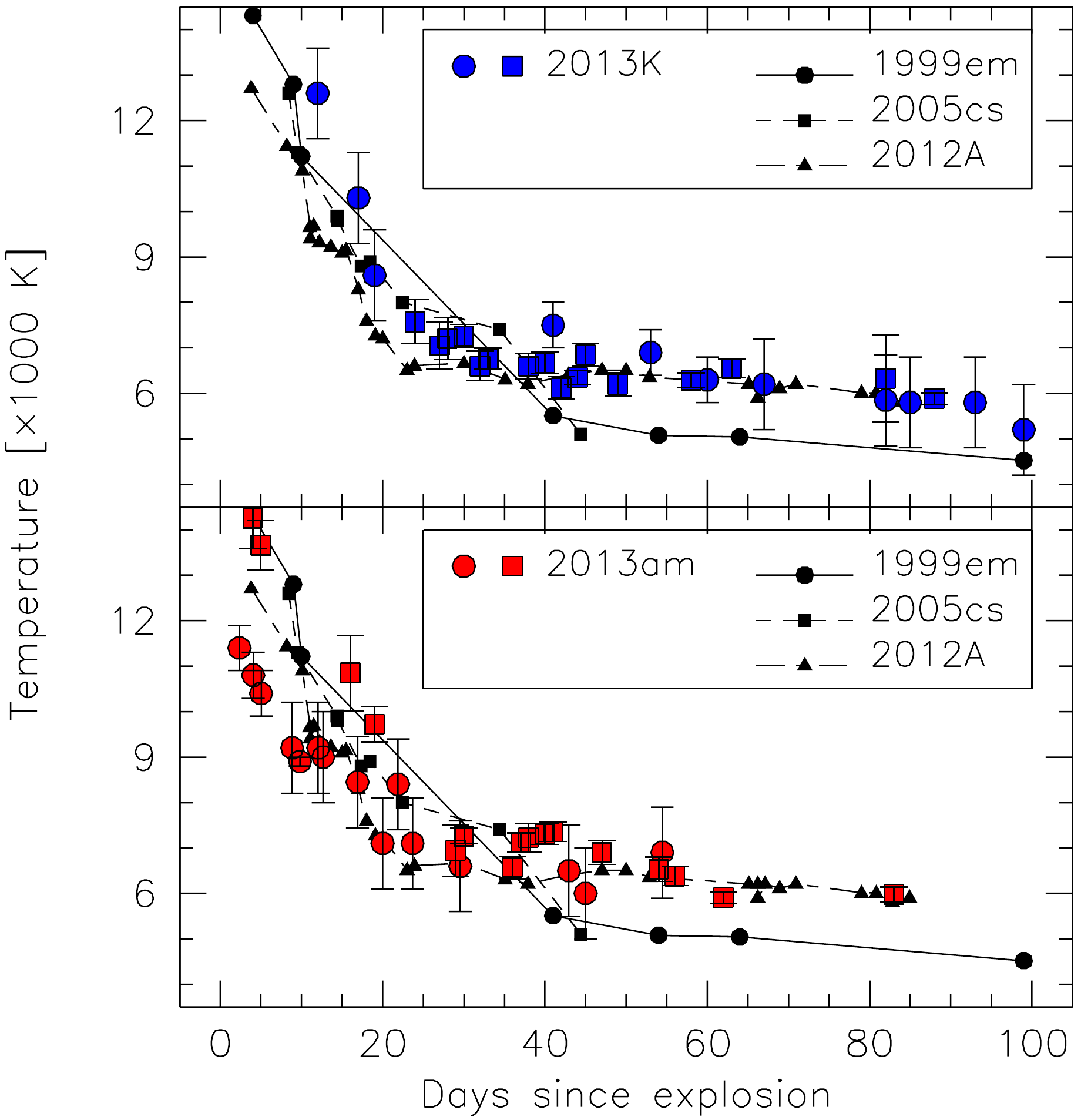}
\caption{Evolution of the continuum temperature of SNe~2013K (top panel) and 2013am (bottom panel), and comparison with SNe~1999em, 2012A and the faint SN~2005cs. For SNe~ 2013K and 2013am, 
the blue and red squares are derived by fitting a black-body to the photometric data, while blue and red circles by fitting a black-body to spectra.}
\label{temp}
\end{figure}

Estimates of the photospheric temperatures are derived from black-body fitting of the spectral continuum 
(the spectra are corrected for the redshift and adopted extinctions), from a few days after explosion up to around two months (using the {\sc iraf/stsdas} task {\sc nfit1d}). Later on the fitting to the continuum becomes difficult, due to both emerging emission lines and increased line blanketing by iron group elements \citep{kasen:2009} which causes a flux deficit at the shorter wavelengths. Actually, when the temperature drops below $10^{4}$ K, the \textit{UV} flux is already suppressed by line blanketing, and the effect becomes even stronger on the blue bands as the temperature decreases down to $\sim8000$ K \citep[][ cf. their Section 3 and Fig. 1]{faran:2018}. Consequently, the black-body fitting to the spectral continuum is performed using the full  wavelength range (typically 3350-10000~\AA\/) solely during the first ten days of evolution. At a later time, wavelengths shorter than 5000~\AA\/ are excluded from the fit because the blanketing affects also the \textit{B}-band. 
We note that estimating the temperature is very challenging, both at early phases, when the peak of the spectral energy distribution is blueward of 3000 \AA\/, i.e. outside our spectral coverage, and at late phases, due to the numerous emission and absorption lines. Setting the sample range of the black-body fitting in order to include or exclude the stronger emission or absorption features, we quantify that the typical uncertainty for the temperature determination is greater than $\sim500$ K.

Also, we estimate the photospheric temperatures by fitting a black-body to the available multi-band photometry (Section~4.1). 
Following \cite{faran:2018}, we exclude from the fitting the bluest bands when the effect of line blanketing on the SED is appreciable.
The uncertainties on the inferred values are estimated with a bootstrap resampling technique, varying randomly the flux of each photometric point according to a normal distribution having variance equal to the statistical error of each point. We do this procedure 1000 times for each epoch, measuring a temperature from each resampling. The error of the temperature is the standard deviations of the new inferred distribution. 

The temperature evolution of SNe~2013K and 2013am is shown in Fig.~\ref{temp}, along with SNe 2012A, 1999em and 2005cs, for comparison. Data obtained by fitting a black-body to the spectra and to the SED are marked with different symbols (circles and squares, respectively), along with the error of the fit. As pointed out by \cite{faran:2018}, at early phases, the {\it Swift} \textit{UV} photometry (available only for SN~2013am) is critical for constraining the black-body fit to the SED, causing differences as large as $\sim2500$ K from the one derived by fitting the coeval spectrum, which covers only wavelengths redward of the \textit{U}-band. After phase +20 d, the deviation is within the error bars. 
For both SNe~2013K and 2013am, the early photospheric temperature is above $1.2 \times 10^{4}$ K and decreases to $\sim 6000$ K within two months, at the end of the plateau phase.

\subsubsection{Expansion velocity}

From each spectrum of SNe~2013K and 2013am, we measure the H${\alpha}$ and H${\beta}$ velocity and, where feasible, the Fe~II $\lambda$ 5169, Sc~II $\lambda$ 5527 and Sc~II $\lambda$ 6246. This is done by fitting Gaussian  profile to the absorption trough in the redshift corrected spectra. 
Following \cite{leonard:2002}, the error estimate includes the uncertainty in the wavelength scale and in the measurement process itself. The wavelength calibration of spectra is checked against stronger night-sky lines using the cross-correlation technique (O~I, Hg~I and Na~I~D, Osterbrock et al. 2000), thus assigning an 1$\sigma$ error of $\sim$ 0.5~\AA\/. To evaluate the fit error, we normalise the absorption feature using the interpolated local continuum fit and we perform multiple measurements adopting different choices for the continuum definition. The standard deviation of the measures is added in quadrature to the uncertainty of the wavelength scale, giving a total, {\it statistical} (not systematic) uncertainty for the velocity of each line ranging from 50 to 200 km~s$^{-1}$ (depending on the signal-to-noise ratio of each spectrum and on the strength of the measured feature). 

As discussed by \cite{dessart:2005}, the velocity measured at the absorption minimum, v$_{abs}$, can overestimate or underestimate the photospheric velocity, $v_{phot}$, especially when considering Balmer absorptions. However, for the Fe~II $\lambda$ 5169 line, the velocity measurement $v_{abs}$ matches $v_{phot}$ to within 5-10\% 
\citep[see ][ their Fig. 14]{dessart:2005}, and also, the Sc~II $\lambda$ 6246 line  is considered a good indicator of the photospheric velocity $v_{phot}$ \citep{maguire:2010}. 

\begin{figure}
\includegraphics[scale=.45,angle=0]{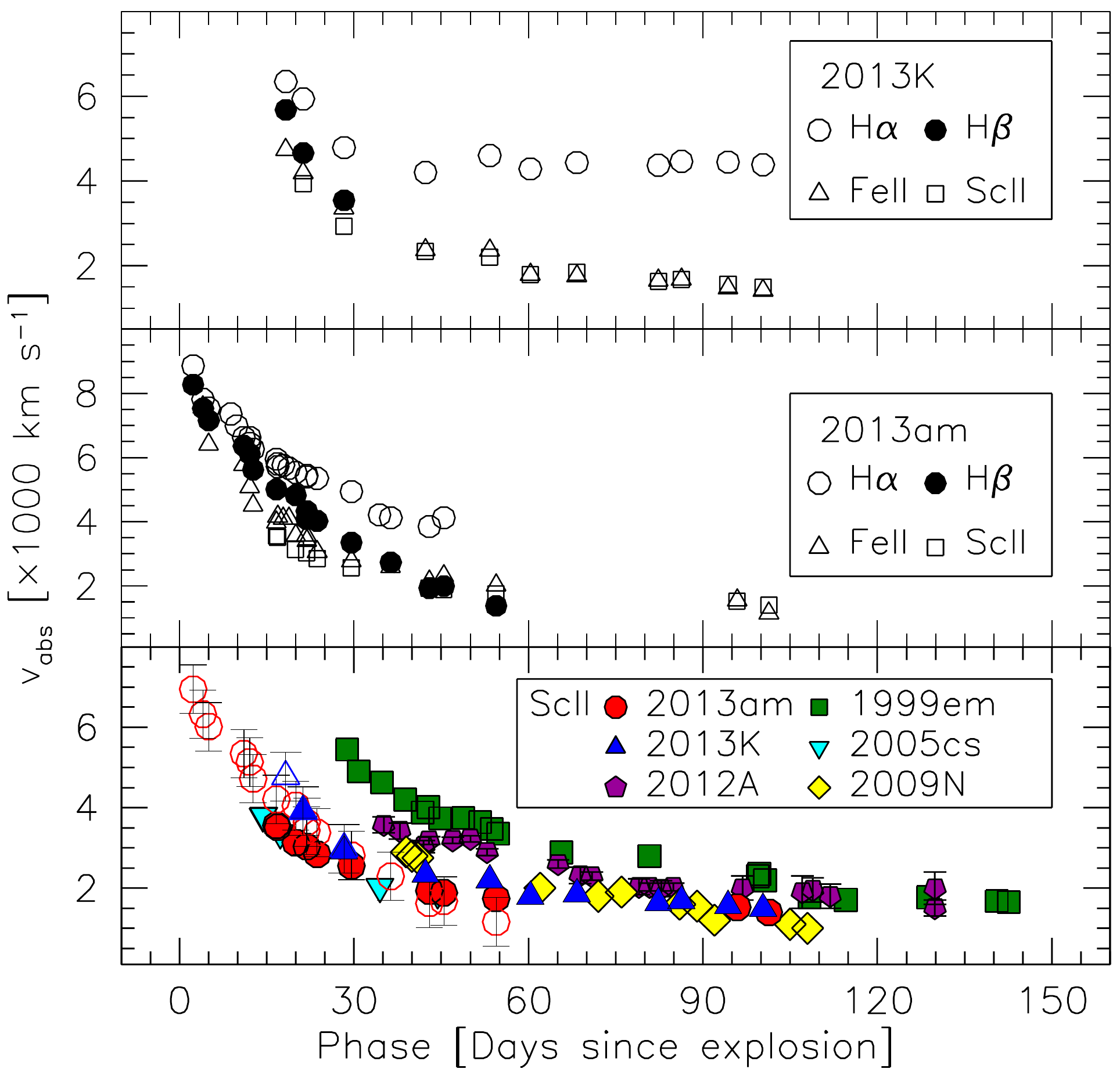}
\caption{ Top panel: evolution of the velocity measured from H$\alpha$, H$\beta$, Fe~II $\lambda$ 5169 and Sc~II $\lambda$ 6246 in SN~2013K and (middle panel) for SN~2013am. The statistical uncertainty in the velocity estimates (maximum 200 km s$^{-1}$) is smaller than the plotted points. Bottom panel: evolution of the velocity of Sc~II $\lambda$ 6246 for SNe~2013K, 2013am, 1999em, 2012A, 2009N and 2005cs. The open (blue) triangles and (red) circles are derived in earlier phases from the absorption minima of H$\beta$ for SN 2013K and 2013am, respectively, applying the linear correlation between H$\beta$ and Fe~II velocities determined by Poznanski et al. 2010. In this case, an additional error of about 300 km s$^{-1}$ is considered and plotted.}
\label{vel}
\end{figure}

In Fig.~\ref{vel} we plot the H${\alpha}$, H${\beta}$ velocity $v_{abs}$, for SN~2013K (top panel) and SN~2013am (middle panel) during the first two weeks of evolution, while the Fe~II and Sc~II features are identified and measured only at later phases. In the bottom panel of Fig.~\ref{vel} we compare $v_{abs} \approx v_{phot}$ of SNe~2013K and 2013am 
as derived from the Sc~II $\lambda$ 6246 line with those of SNe 1999em, 2012A, 2009N and 2005cs. At early phases, we consider the H$\beta$ velocity as a proxy to Fe~II $\lambda$ 5169, applying the linear correlation derived by \cite{poznanski:2010}, i.e. $v_{FeII} = 0.84 \pm 0.05$~$v_{H\beta}$, and plotting these data as open (blue) triangles and (red) circles, respectively. 
At phase $\sim95$ d, the velocity settles around 1500~km~s$^{-1}$ for both SNe~2013K and 2013am. 

\subsubsection{Nebular spectra}

\begin{figure*}
\includegraphics[scale=0.5,angle=0]{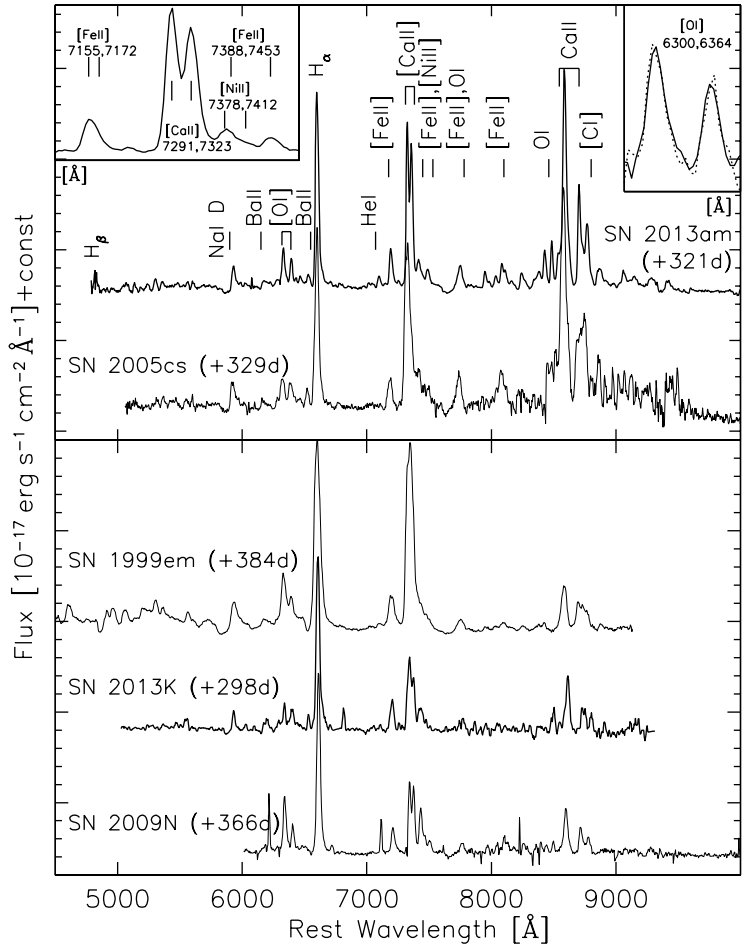}
\caption{Top panel: comparison between nebular spectra, obtained about 1 yr after explosion, of SN~2013am and SN~2005cs, and lines identification. Bottom panel: nebular spectra of SN~2013K, SN~1999em and SN~2009N. The inset on top left corner  shows a closed up view of the SN~2013am spectrum centred on [Ca~II] $\lambda\lambda$ 7291, 7323; the positions of [Fe~II] $\lambda$ 7155, [Fe~II] $\lambda$ 7172, [Fe~II] $\lambda$ 7388, [Fe~II] $\lambda$ 7453, 
[Ni~II] $\lambda$ 7378 and [Ni~II] $\lambda$ 7412 are also marked. The inset on top right corner shows the [O~I] $\lambda\lambda$ 6300, 6364 doublet for SN~2013am at +320.7~d (solid line), compared with the spectrum at phase +280.9~d (dotted line). }  
\label{fig_neb}
\end{figure*}

\begin{figure}
\includegraphics[scale=.44,angle=0]{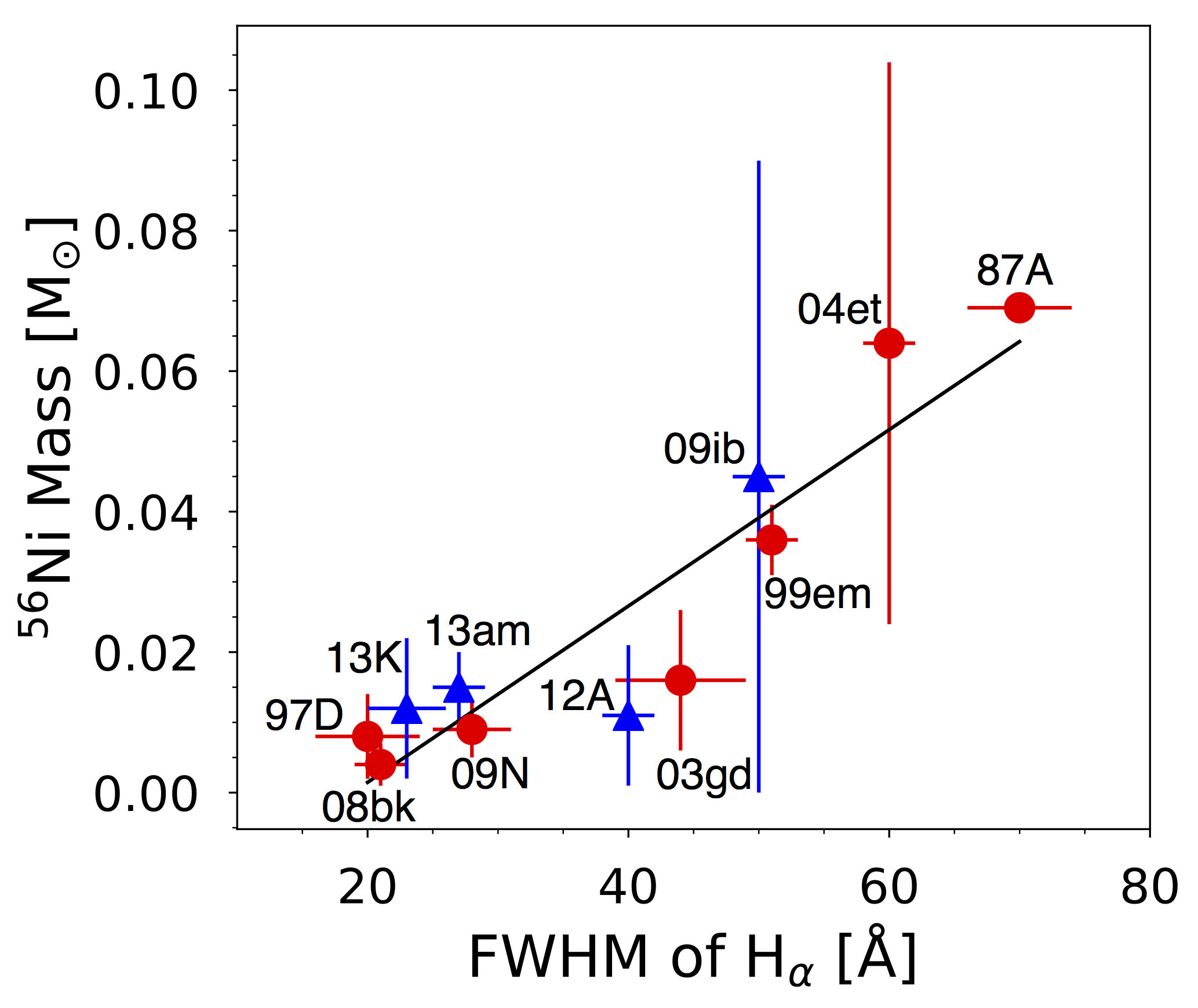}
\caption{Correlation between the FWHM of H${\alpha}$ (corrected for instrumental resolution) obtained from nebular spectra of a sample of well studied SNe and the ejected mass of $^{56}$Ni obtained from the nebular tail of their light curves (new data for SNe~2009ib, 2012A, 2013K and 2013am are plotted as blue triangles). See text for details of the derived weighted least squares fit to the data.} 
\label{fwhm}
\end{figure}

In Fig.~\ref{fig_neb} we compare the nebular spectra of SNe~2013K and 2013am, taken around 1 year after explosion, with the standard Type~IIP SN~1999em \citep{elmhamdi:2003}, the faint SN~2005cs \citep{pastorello:2009} and the intermediate-luminosity SN~2009N \citep{takats:2014}. 
The narrow H${\alpha}$ emission lines have FWHM (corrected for the instrumental resolution) of about $27\pm 2$ \AA\/ for SN~2013am and $23\pm3$ \AA\/ for SN~2013K, corresponding to velocities of $\sim1200$ and $\sim1000$ km s$^{-1}$, respectively. Maguire et al. 2012 (see their Sections 4.1 and 7) derived an empirical relation between the mass of $^{56}$Ni and the corrected FWHM of H$\alpha$ for a sample of seven SNe, from the underluminous SN 1997D to SN 1987A. Including SNe~2013am, 2013K (cf. Section~\ref{bolometric}) and SN~2012A \citep[FWHM = $40 \pm 2$~\AA\,, M($^{56}$Ni) = 0.011 M$_\odot$, see][]{tomasella:2013}, we double the sample size in the sub-energetic tail. Additionally, we include the intermediate-luminosity  SN~2009ib \citep[FWHM = $50 \pm 2$~\AA\,, M($^{56}$Ni) = 0.046 M$_\odot$, see][]{takats:2015} with moderate expansion velocities, as already done for Fig.~\ref{nugent}. 
The updated plot of FHWM versus ejected $^{56}$Ni mass is shown in Fig.~\ref{fwhm}. We performed a linear least squares fit to the data, weighting each point by its uncertainties, and find that 

\begin{equation}
\mathrm{M}(^{56}\mathrm{Ni}) = (1.25 \pm 0.23)  \times 10^{(-0.024 \pm 0.009) \times \mathrm{FWHM}} \times 10^{-3} \mathrm{M}_\odot
\end{equation}

with $\sigma = 0.009 \pm 0.012$.
Pearson's correlation coefficient and Spearman's rank correlation are 0.921 and 0.909, respectively.

The Na~I D (still showing residual P-Cygni absorption) and Ca~II NIR triplet lines are well detected. The feature at around 7300 \AA\ that is always observed in the nebular spectra of Type IIP SNe, is identified as the [Ca~II] $\lambda\lambda$ 7392, 7324 doublet. The individual components of the [Ca~II] doublet are resolved. In the last spectrum of SN~2013am, the H${\alpha}$ and [Ca~II] doublet have comparable luminosities, resembling the faint SN~2005cs (cf. Fig.~\ref{fig_neb}) rather than normal or intermediate-luminosity Type~IIP SNe. In both SNe~2013K and 2013am, the [O~I] $\lambda\lambda$ 6300, 6364 doublet is clearly detected, though much weaker than [Ca~II].  
Several lines of [Fe~II] \citep[multiplets 19 and 14, with the contribution of multiplet 30, cf.][]{benetti:2001} are visible in the nebular spectra, along with other weaker features that can be attributed to [Fe~I], Fe~I, Fe~II, O~I and Ba~II. The latter is clearly identified as the 6497 \AA\/ line blueward of H${\alpha}$. Ba~II $\lambda\lambda$ 5854, 6497 (the first component is blended with Na~I D) together with Ba~II $\lambda$ 6142 (blended with Fe~I, Fe~II) were previously identified in underluminous Type IIP SNe, such as SNe~1997D \citep{turatto:1998}, 2005cs \citep{pastorello:2009}, and 2008bk \citep{lisakov:2017}, but also in the intermediate-luminosity Type~IIP SNe~2008in \citep{roy:2011}, 2009N \citep{takats:2014}  and 1999em \citep{leonard:2002}. The appearance of relatively strong Ba~II lines is likely due to the combination of a low temperature (below $\sim$ 6000 K, see Hatano et al. 1999; Turatto et al. 1998) and low expansion velocity (i.e. narrow, unblended lines are better seen, while in standard SNe II the higher expansion rate at the base of the H-rich envelope causes the contributions of Ba II 6497 \AA\/ and H$\alpha$ to merge into a single spectral feature, as discussed by Lisakov et al. 2017). 
We note the presence in SN~2013am of a feature redward of 7000 \AA\/, which could be identified with He~I $\lambda$ 7065, even if it is rarely seen in nebular spectra (detected in SN~2008bk by Maguire et al. 2012) or with another metallic forbidden line. In both SNe, we can identify [C~I] $\lambda$~8727 which is a helium burning ash and a tracer of the O/C zone.

The ratio ($\Re$) between the luminosities of the [Ca~II] $\lambda\lambda$ 7392, 7324  and [O~I] $\lambda\lambda$ 6300, 6364 doublets appears to be almost constant at late epochs, and has been proposed as a diagnostic for the core mass and, consequently, for the main-sequence mass ($M_{\rm ZAMS}$) of the SN progenitor \citep[cf.][ see also Jerkstrand et al. 2012]{fransson:1987,fransson:1989,woosley:1995}.  $\Re$ is inversely proportional to the precursor's $M_{\rm ZAMS}$. From the emission line spectra of the peculiar Type~II SN 1987A it was found $\Re \approx 3$ \citep[][]{li:1992,li:1993}. Extensive studies on the blue progenitor of this event indicate an initial mass in the range of $14-20$ M$_{\odot}$, i.e. close to the suggested upper limit of 19 M$_{\odot}$ for the progenitor mass of Type IIP SN \citep[][ see also the reviews by Arnett et al. 1989; Smartt 2009]{dwarkadas:2014,smartt:2015b}. On the opposite side, the nebular spectrum of SN~2005cs yields a high value of this luminosity ratio, $\Re \approx4.2 \pm 0.6$ \citep[cf.][]{pastorello:2009}, which is consistent with a small He core and hence a low main-sequence mass for the progenitor, close to the minimum initial mass that can produce a SN, namely 8$\pm1$~M$_{\odot}$ \citep{smartt:2009}. Indeed, pre-explosion HST imaging revealed a $M_{\rm ZAMS} \sim 7-13$~M$_{\odot}$ RSG as the precursor of SN~2005cs \citep[cf.][ and references within]{eldridge:2007}. Late-time HST observations by \cite{maund:2014} confirm the progenitor identification of SN~2005cs, and the progenitor mass was refined to $M_{\rm ZAMS} = 9.5^{+3.4}_{-2.2}$ M$_{\odot}$. In between, we find that the nebular spectrum of SN~1999em plotted in Fig.~\ref{fig_neb} is characterised by $\Re \approx 3.6 \pm 0.5$, suggesting a progenitor of moderate main-sequence mass, in the range of $12-15$~M$_{\odot}$. Coherently, for this event $M_{\rm ZAMS} \sim 12 \pm 1$ M$_{\odot}$ was derived by \cite{smartt:2002}, analysing the pre-SN images of the Canada-France-Hawaii Telescope archive, while the hydrodynamical modelling in \cite{elmhamdi:2003} suggested  $M_{\rm ZAMS} \sim 12-14$ M$_{\odot}$. 

Similar to SN~1999em, the line strength ratio from the spectrum of SN~2013am at phase +320.7 d, is  $\Re \approx 3.7 \pm 0.5$; an analogous value is derived for SN~2013K at $+$298.2 d, although with a larger uncertainty due to the lower signal-to-noise ratio of the spectrum. Again, these intermediate $\Re$ are suggestive of moderate-mass progenitors ($12-15$ ~M$_{\odot}$). The hydrodynamical modelling described in Section~\ref{hydro} supports this hint. We caution however that the line ratio is only a rough diagnostic of the core mass, due to the contribution to the emission from primordial O in the hydrogen zone \citep{maguire:2012,jerkstrand:2012}. Moreover, mixing makes it difficult to derive relative abundances from line strengths, as discussed in \cite{fransson:1989}. An example of these effects can be seen for SN~2012A, were we find $\Re \approx 2$ (cf. Tomasella et al. 2013, their table 8), which would suggest a higher mass than that inferred by \cite{tomasella:2013} either through the direct detection of the progenitor in pre-SN images ($10.5^{+4.5}_{-2}$~M$_{\odot}$), or hydrodynamical modelling of the explosion ($14-15$ M$_{\odot}$).

\subsubsection{Nickel and Iron forbidden lines}\label{ni_ir}

SNe~2013K and 2013am show distinct, narrow [Fe~II] $\lambda$ 7155 and [Ni~II] $\lambda$ 7378 lines in their nebular spectra (fig.~\ref{fig_neb}). The emissivity ratio of [Ni~II] to [Fe~II] ($L_{7378}/L_{7155}$) is also considered a diagnostic of the Ni to Fe ratio, as discussed by Jerkstrand et al. (2015, 2015a). From a Gaussian fit to these lines, we determine a ratio $\sim 1$ for both the SNe. This implies an abundance number ratio Ni/Fe~$\approx 0.06 \pm 0.01$, i.e similar to the solar value  \citep[][ their Section~5.1.4]{jerkstrand:2015}.
According to \cite{jerkstrand2:2015}, SNe that produce solar or subsolar Ni/Fe ratios must have burnt and ejected only Oxygen-shell material. On the other hand, a larger Ni/Fe ratio would imply that there was explosive burning and ejection of the silicon layer, as in the case of SNe~2006aj and 2012ec (see Jerkstrand et al. 2015, Maeda et al. 2007 and Mazzali et al. 2007). The solar Ni/Fe ratio measured for both SNe~2013K and 2013am again would suggest medium-mass progenitors, with M$_\mathrm{ZAMS}$ of about 15 M$_{\odot}$ (cf. Jerkstrand et al. 2015a, their Fig.~8).

\subsubsection{Oxygen forbidden lines}

\cite{jerkstrand:2012} used their spectral synthesis code to model the nebular emission line fluxes for SN explosions of various progenitors, and find the [O I], Na I D and Mg I] lines to be the most sensitive to ZAMS mass. 
In particular,  the [O I] lines appear to be a reliable mass indicator \citep[see also ][]{chugai:1994,chugai:2000,kitaura:2006}. Unfortunately,  the nebular spectrum of SN~2013K has a low signal-to-noise ratio, and there appears to be significant contamination from the Fe~I $\lambda$ 6361 line \citep[][ see their fig.~17]{dessart:2011}. The [O~I] line flux relative to $^{56}$Co luminosity (assuming M($^{56}$Ni) = $0.012\pm0.010$ M$_{\odot}$ and a distance of 34 Mpc) is estimated to be of the order of $3-4$ per cent. Based on the modelling of Jerkstrand et al. 2012 (their Fig.~8), this ratio would suggest a moderately massive star, between 15 and 19 M$_{\odot}$, as the precursor of SN~2013K.
Conversely, from the high signal-to-noise nebular spectrum of SN~2013am taken at phase +320.7 d (assuming M($^{56}$Ni) = $0.015\pm0.006$ M$_{\odot}$ and a distance of 12.8 Mpc) we find that 
the [O I] $\lambda\lambda$ 6300, 6364 flux relative to $^{56}$Co luminosity is only $\sim0.6$  per cent. The low value seems to point to a progenitor for SN~2013am of less than 12~M$_{\odot}$. This is in contrast with other indicators,  including the luminosity ratios [Ca~II]$/$[O~I] and [Ni~II]$/$[Fe~II] calculated previously, and the hydrodynamical model (cf. Section~\ref{hydro}), which indicate for both SNe moderate-mass progenitors, in the range $12-15$ M$_{\odot}$.

We caution that the nebular models in \cite{jerkstrand:2012} are computed for a higher ejected $^{56}$Ni mass (M($^{56}$Ni) = 0.062 M$_{\odot}$, as derived for SN 2004et) and higher velocities of the core material ($\sim 1800$ km s$^{-1}$, which is at least 30 per cent larger than the velocities estimated for SNe~2013K and 2013am). 
\cite{jerkstrand:2015} investigated the sensitivity of line flux ratios by varying the mass of ejected $^{56}$Ni from 0.06 to 0.03 M$_{\odot}$, and concluded that the diagnostics are reliable in this range. However, SNe~2013K and 2013am have much smaller M($^{56}$Ni). New models for both low-velocity and low M($^{56}$Ni) Type~II SNe are being produced (Jerkstrand et al., in prep.), and detailed comparison of the new models with the spectra of SNe~2013K and 2013am and other Type~II SNe will be included in this work.

Simulations of stellar collapse and low energy explosions performed by \cite{kitaura:2006} point to a clear difference between low-mass and more massive SN progenitors. The former, like SN 1997D (Chugai \& Utrobin 2000), eject a very small amount of O (only a few $10^{-3}$ M$_{\odot}$), whereas the latter produce up to a solar mass or more \citep[in the range $1.5-2$ M$_{\odot}$ for SN~1987A, see][]{chugai:1994}. Rescaling the luminosity of [O~I] doublet and the M($^{56}$Ni) to SN~1987A, as done by \cite{elmhamdi:2003} for SN~1999em (their Section~3 and eq.~1), a very rough estimate of the ejected O mass in SN~2013am is $\sim 0.1-0.2$ M$_{\odot}$. 

Finally, the [O~I] $\lambda\lambda$ 6300, 6364 line flux ratio can also be used to determine the optical depth of those transitions \citep{spyromilio:1991,spyro:1991,spy:1991,li:1992}, and to calculate the number density of neutral Oxygen in the ejecta. The flux ratio $F_{6364}/F_{6300}$ is around unity in the optically thick limit. As the SN ejecta expand, the ratio decreases to about $1/3$ in the optically thin limit. For SN~2013am, we find $F_{6364}/F_{6300} \approx 0.65$ at phase +280.9 d, decreasing to $\approx 0.56$ at phase +320.7 d, marking the incomplete transition from the optically thick towards the optically thin case. This is similar to what was found for a sample of Type IIP SNe by Maguire et al. (2012; cf. their Section 5.4 and fig.~12).  Following \cite{spy:1991}, it is possible to uniquely determine the number density of neutral O with only one observation between the optically thick and optically thin limits. For SN~2013am, we derive a neutral O number density of about $\sim 1.0\pm0.5 \times10^9$ cm$^{-3}$.

\subsection{Infrared spectra}

\begin{figure}
\includegraphics[scale=0.43,angle=0]{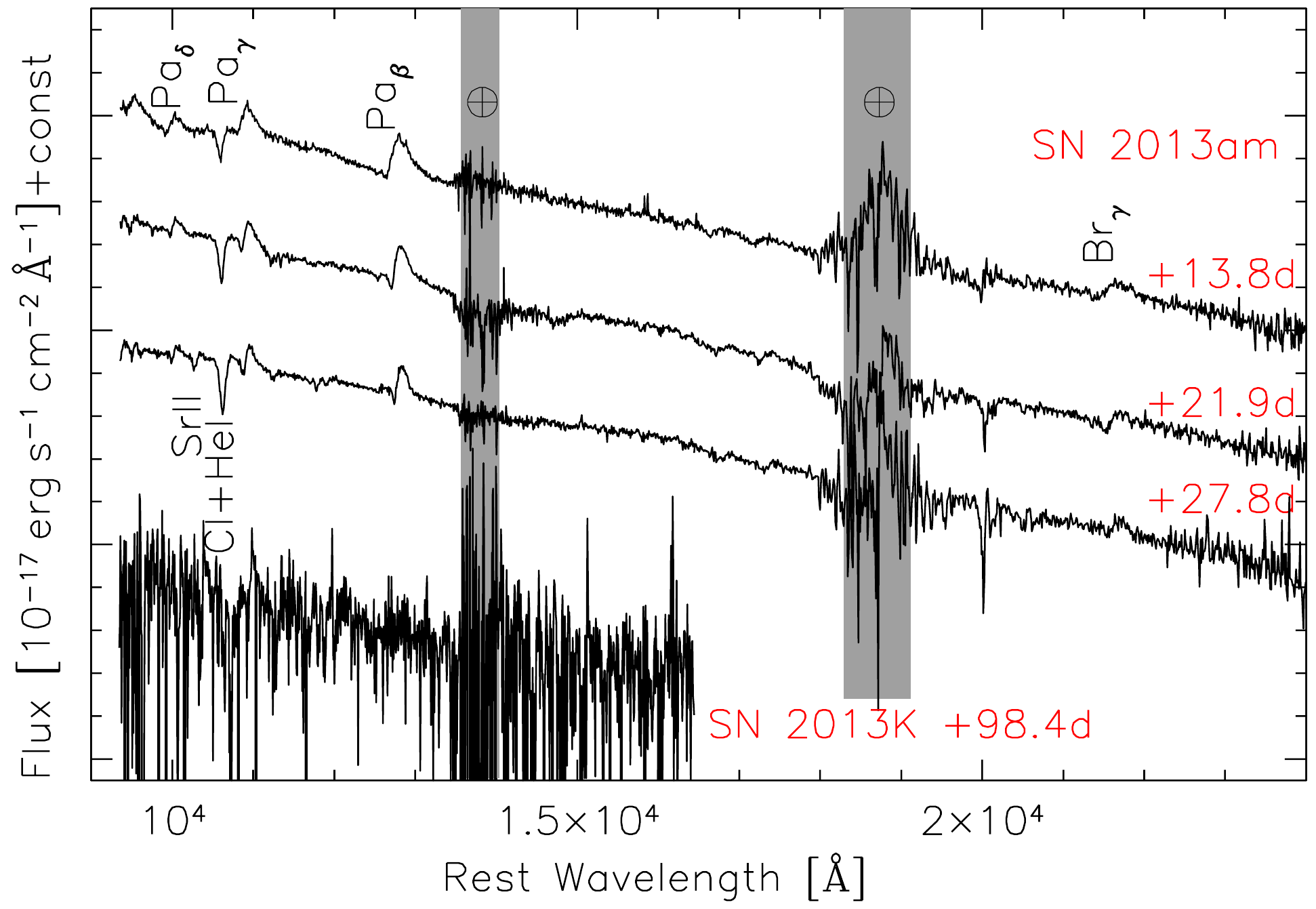}
\caption{Sequence of NIR spectra of SN 2013am taken with ESO NTT$+$SOFI at phases +13.8, +21.9, +27.8~d (top), shown along with the spectrum of SN~2013K at +98.4~d (bottom). Paschen series, Br${\gamma}$, the blend of C~I $\lambda$ 10691 and He~I $\lambda$ 10830, and Sr~II $\lambda$ 10327 lines are identified. The positions of major telluric features are marked with $\oplus$ symbols. }  
\label{fig_nir}
\end{figure}

NIR spectra were collected for SN~2013am (three epochs) and SN~2013K (one epoch) during the photospheric phase with NTT$+$SOFI (cf. Tables~\ref{telescope_spec} and \ref{telescope_spec2}). They are plotted in Fig.~\ref{fig_nir}. The Paschen series (with the exclusion of Pa${\alpha}$, which is contaminated by a strong telluric band), Br${\gamma}$, the blend of C~I $\lambda$ 10691 and He~I $\lambda$ 10830, and Sr~II $\lambda$ 10327 are identified. In the spectra of SN~2013am, a relatively strong feature redward of Pa${\gamma}$ can be identified as the O~I $\lambda$ 11290 Bowen resonance fluorescence line \citep{pozzo:2006}. 

For SN~2013am, the expansion velocity of ejecta, as measured from the absorption minima of Pa${\beta}$ and Pa${\gamma}$, is $\sim 5100$~km~s$^{-1}$ at phase +13.8 d, decreasing to $\sim 3400$~km~s$^{-1}$  (+21.9 d) and $\sim 2500$~km~s$^{-1}$ after one month of evolution. As expected, this expansion velocity is lower than that obtained at similar epochs from the Balmer lines. Rather, it is comparable to the values obtained from the metal lines in the optical spectra (cf. Fig.~\ref{vel}). After smoothing the noisy NIR spectrum of SN~2013K at phase $+98.4$ d, an expansion velocity from Pa${\gamma}$ of about $\sim 2500 - 3000$~km~s$^{-1}$ is found, which lies in between the velocities derived from the H${\alpha}$ absorption minimum and Sc~II $\lambda$ 6246 at the same epoch.

\section{Hydrodynamical modelling}\label{hydro}

The physical properties of the SN~2013K and SN~2013am progenitors at the time of the explosion, namely the ejected mass ($M_{ej}$), the initial radius ($R$), and the total explosion energy ($E$), are derived from the main observables (i.e. the pseudo-bolometric light curve, the evolution of line velocities and the photospheric temperature), using a well-tested radiation-hydrodynamical modelling procedure.\footnote{This hydrodynamical modelling was previously applied to other Type~II SNe, e.g. SNe~2007od, 2009bw, 2009E, 2012A, 2012aw, 2012ec and 2013ab; (see Inserra et al. 2011, 2012; Pastorello et al. 2012; Tomasella et al. 2013; Dall'Ora et al. 2014; Barbarino et al. 2015; Bose et al. 2015, respectively).}
A complete description of this procedure is available in \cite{pumo:2017}.  To obtain the best fit, a two step procedure is adopted using two different codes: 
1) a semi-analytic code \citep{zampieri:2003}, which solves the energy balance equation for ejecta of constant density in homologous expansion; and 2) a general-relativistic, radiation-hydrodynamics Lagrangian code 
\citep{pumo:2010,pumo:2011}, which was specifically tailored to simulate the evolution of the physical properties of the SN ejecta and the behaviour of the main SN observables up to the nebular stage, taking into account both 
the gravitational effects of the compact remnant and the heating due to the decay of radioactive isotopes synthesised during the explosion. The former code is used for a preliminary analysis aimed at constraining the parameter space describing the SN progenitor at the time of the explosion and, consequently, to guide more realistic, but time consuming simulations performed with the latter code.

\begin{figure*}
\includegraphics[scale=.5,angle=0]{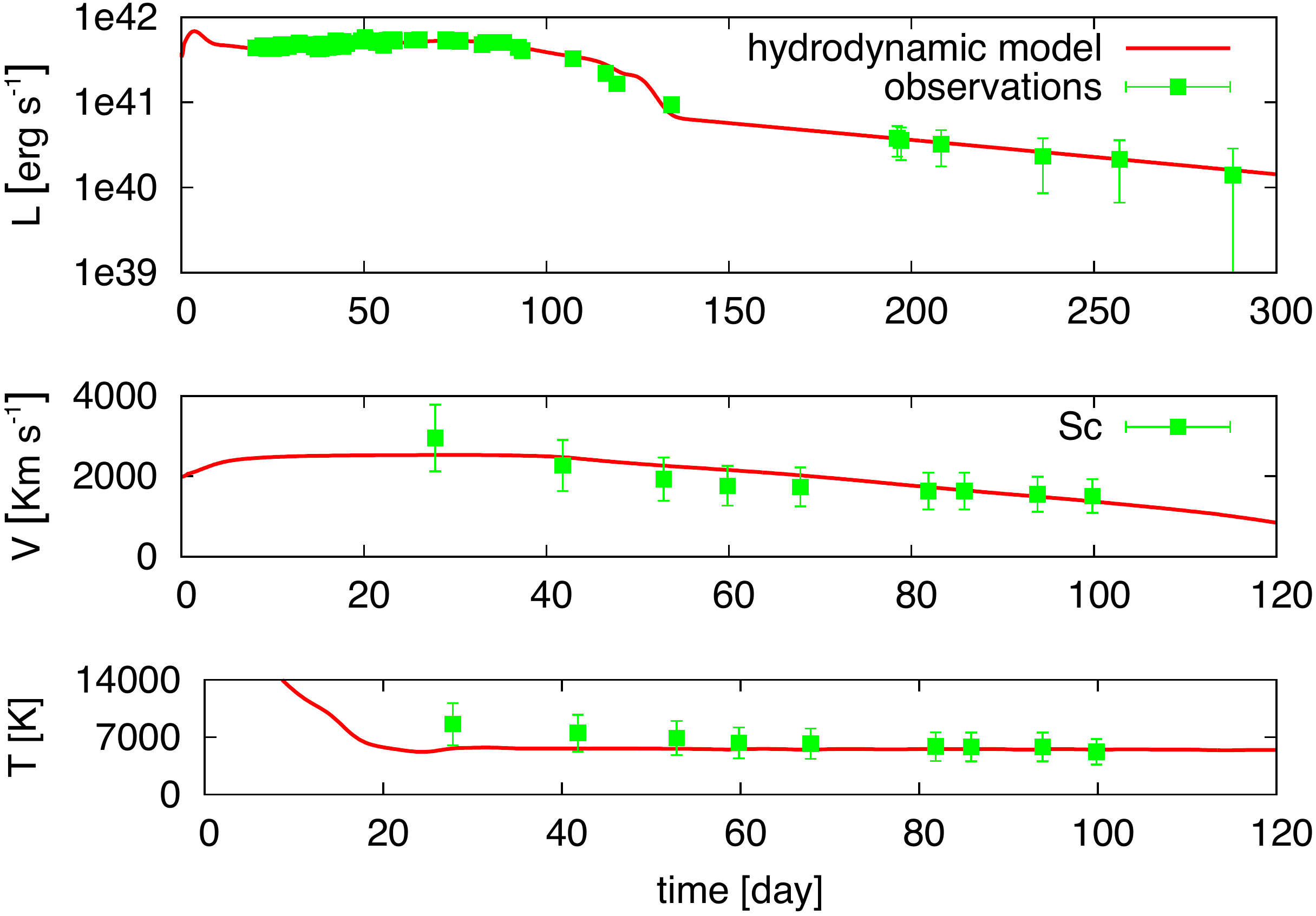}
\caption{Comparison of the evolution of the main observables of SN 2013K with the best-fit model computed with the general-relativistic, radiation-hydrodynamics code. The best-fit model parameters are: $E = 0.34\pm0.10$ foe, 
$R = 3.2\pm0.5 \times 10^{13}$ cm, $M_{ej} = 12.0\pm1.8$ M$_{\odot}$. Top, middle, and bottom panels show the bolometric light curve, the photospheric velocity, and the photospheric temperature as a function of time. To estimate the photospheric velocity from the observations, we use the minima of the profile of the Sc~II lines.} 
\label{fig:model_13K}
\end{figure*}

\begin{figure*}
\includegraphics[scale=.5,angle=0]{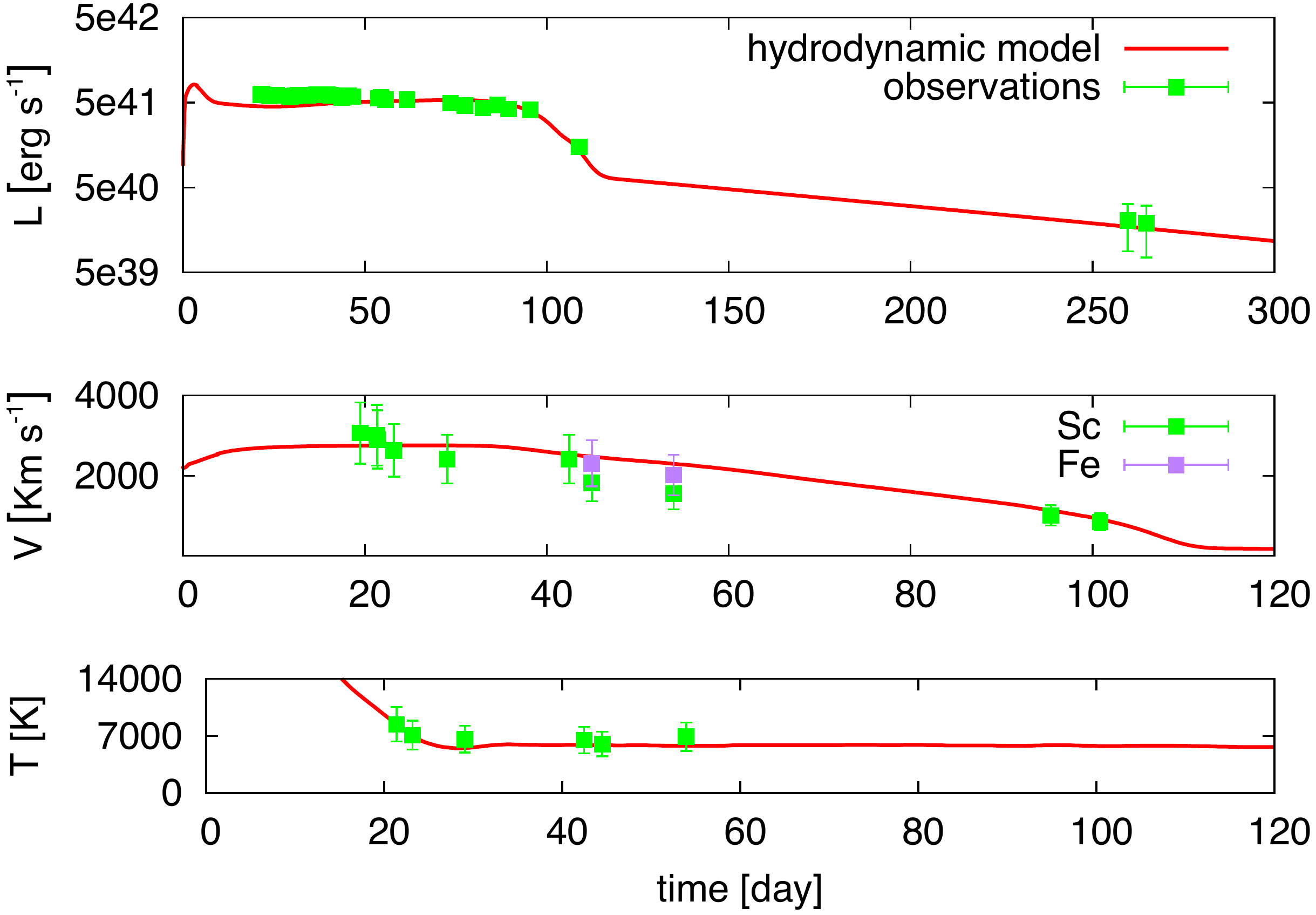}
\caption{Same as Fig.~\ref{fig:model_13K}, but for SN 2013am. The best-fit model parameters are: $ E = 0.40\pm0.12$ foe, $ R = 2.5\pm0.4 \times 10^{13}$ cm, $M_{ej} = 11.5\pm1.7$ M$_{\odot}$. In the middle panel, Fe~II velocity at phase $+45.4$ and $+54.4$ d is also plotted, as comparison.} 
\label{fig:model_13am}
\end{figure*}

Adopting the explosion epochs of Section~\ref{2}, and the bolometric luminosities and M($^{56}$Ni) as in Section~\ref{bolometric}, we compute the best-fitting models, shown in Figs~\ref{fig:model_13K} and \ref{fig:model_13am}, for SNe 2013K and 2013am, respectively. We note the general agreement, within the errors, of the best-fitting models with the observables, unless for the Sc~II line velocities of SN~2013am at phase $+45.4$ and $+54.4$ d; instead, at these two epochs, the model fit well the velocity of Fe~II (cf. Figure~\ref{fig:model_13am}, middle panel). 

The best-fit model of SN 2013K has $E = 0.34$ foe, $R = 3.2 \times 10^{13}$ cm ($\sim460$ R$_{\odot}$), and $M_{ej} = 12$ M$_{\odot}$. Adding the mass of the compact remnant ($\sim 1.3-2$ M$_{\odot}$) to that of the ejected material, we obtain a total stellar mass of $\sim 13.3-14$ M$_{\odot}$ at the point of explosion. 
Concerning SN~2013am, the best-fit model has $ E = 0.40$ foe, $R = 2.5 \times 10^{13}$ cm ($\sim360$~R$_{\odot}$), and $M_{ej} = 11.5$ M$_{\odot}$, resulting in a total stellar mass of $\sim12.8-13.5$ M$_{\odot}$ at explosion. 
We estimate that the typical error due to the $\chi^{2}$ fitting procedure is about $15$ per cent for $M_{ej}$ and $R$, and $30$ per cent for $E$. These errors are the $2\sigma$ confidence intervals for one parameter based on the $\chi^{2}$ distributions produced by the semi-analytical models.
For both SNe~2013K and 2013am, the outcomes of modelling are consistent with low-energy explosions of moderate-mass red supergiant stars.

\section{Summary and further comments}

\begin{figure}
\includegraphics[scale=.35,angle=0]{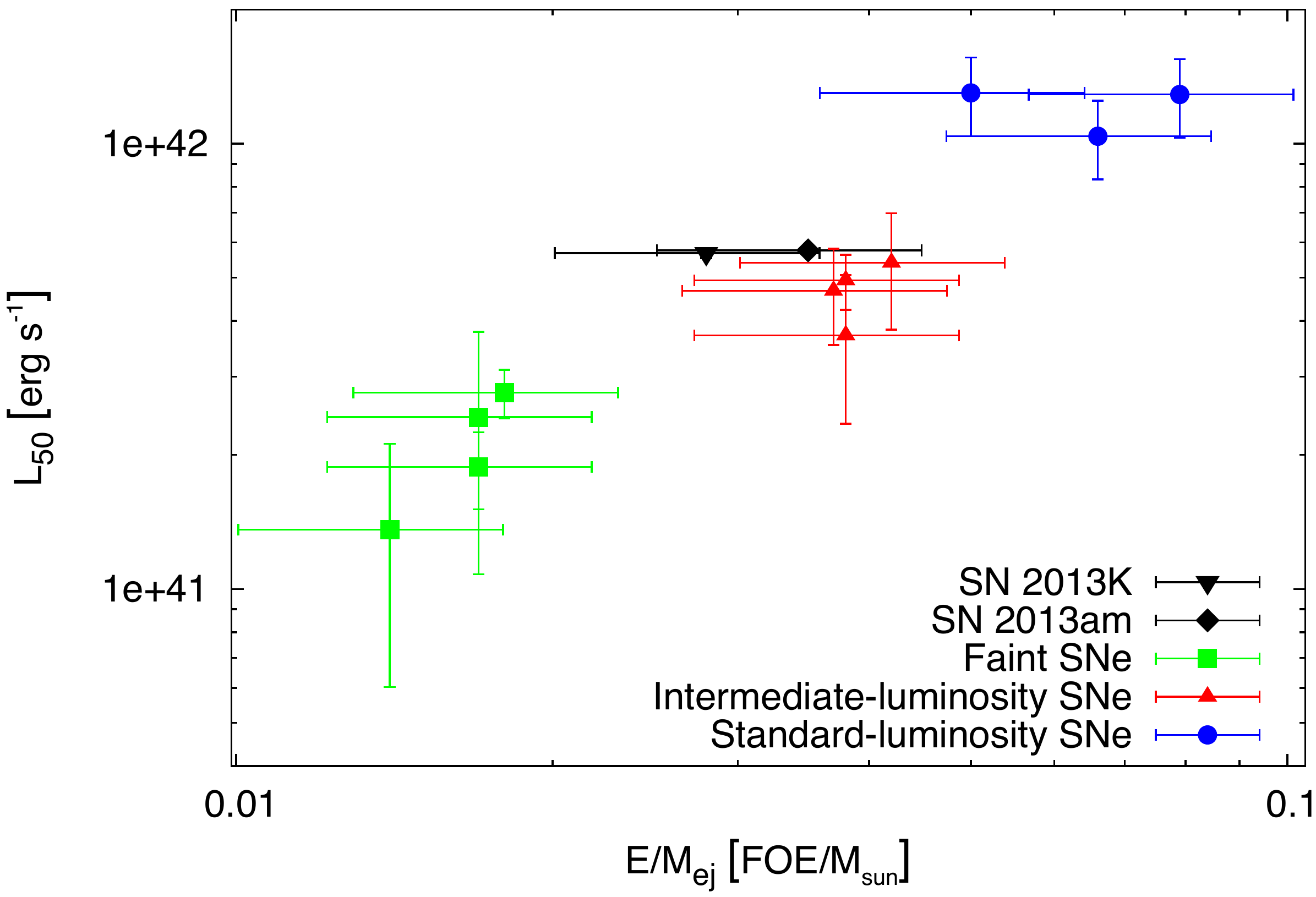}
\includegraphics[scale=.35,angle=0]{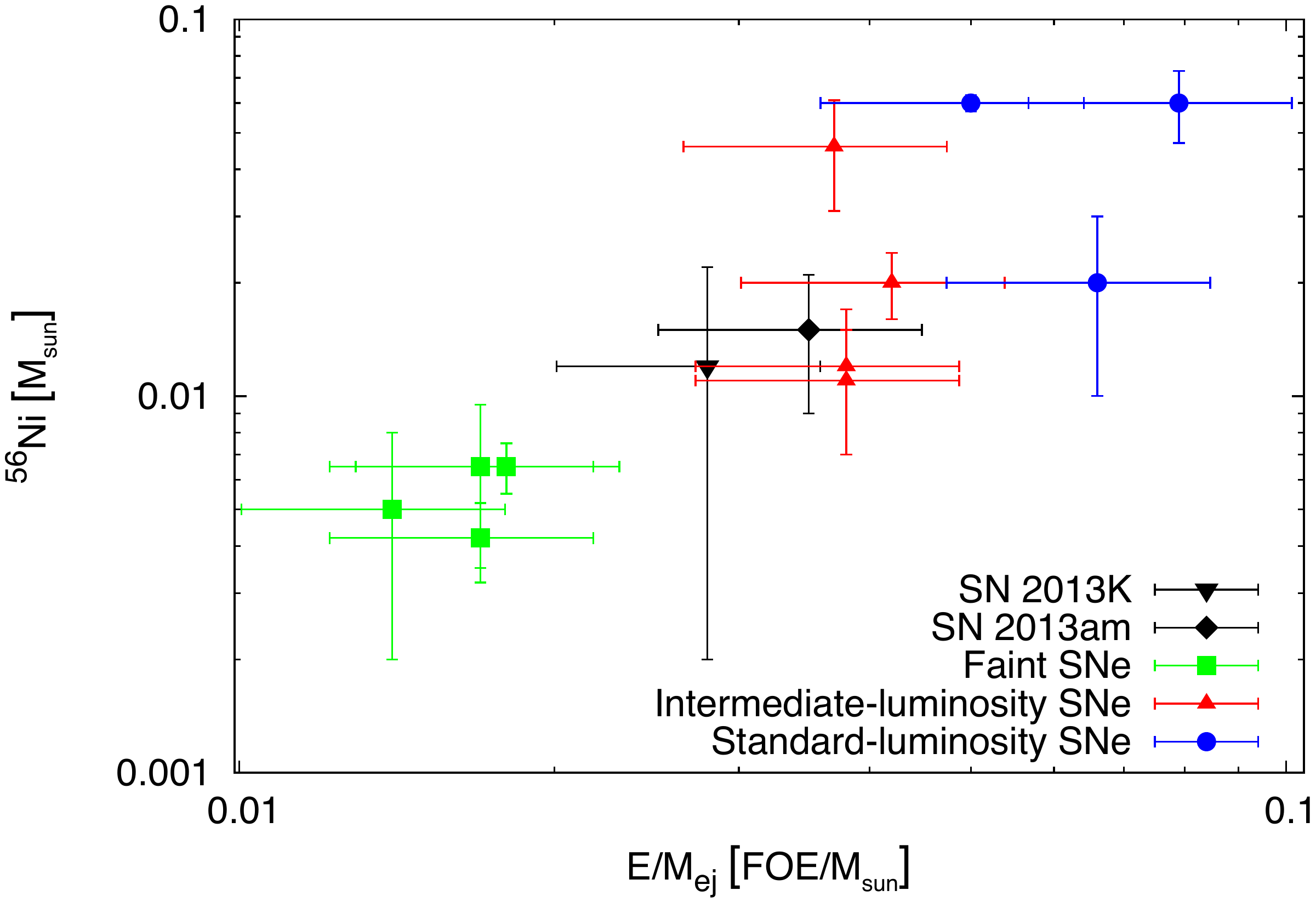}
\caption{Correlations between the plateau luminosity (top) and $^{56}$Ni mass (bottom) with the $E/M_{ej}$ ratio, as in Pumo et al. 2017 (their table~2, figs~5, 6), including SNe~2013K and 2013am (black symbols). Like in Pumo et al. 2017 (to which we refers for details), the errorbars on the $E/M_{ej}$ ratios are estimated by propagating the uncertainties on $E$ and $M_{ej}$, adopting a value  of 30\% for the relative errors of $E$ and 15\% for that of $M_{ej}$. } 
\label{distribution}
\end{figure}

We collected optical and NIR observations of SNe~2013K and 2013am. From the photospheric to the nebular phases, the spectra of these events show narrow features, indicating low expansion velocities ($\sim 1000 -1500$ km s$^{-1}$ at the end of the plateau), as found in the sub-luminous SN~2005cs. In the photospheric phase, we identify features arising from Ba~II, which are typically seen in the spectra of faint Type~IIP SNe. Futhermore, the emission line ratios in the nebular spectra of SN~2013am resemble those of SN~2005cs. The NIR spectra show the Paschen and Bracket series, along with He~I, Sr~II, C~I and Mg~I features, typical of SNe~IIP.

The bolometric luminosities of SNe~2013K and 2013am  ($\sim 1.5 \times 10^{41}$ erg s$^{-1}$ at the plateau), 
are intermediate between those of the underluminous and normal Type~IIP SNe. Indeed, the ejected mass of $^{56}$Ni estimated from the radioactive tail of the bolometric light curves, is $0.012\pm0.010$ 
and $\sim0.015\pm0.006$~M$_{\odot}$ for SN~2013K and SN~2013am respectively: twice the amount synthesised by the faint SNe~1997D or 2005cs, but 3 to 10 times less than that produced by normal SNe IIP events. Similar ejected $^{56}$Ni masses were derived for SNe~2012A, 2008in and 2009N \citep{tomasella:2013,roy:2011,takats:2014}.

We used radiation-hydrodynamics modelling of observables \citep{pumo:2011} to derive the physical properties of the progenitors at the point of explosion for SNe~2013K and 2013am, finding $M_{ej}$ ($12\pm1.8$~M$_{\odot} $\textit{vs.} 
$11.5\pm1.7$ M$_{\odot}$, respectively), $R$ ($3.2\pm0.5 \times 10^{13}$~cm \textit{vs.} $2.5\pm0.4 \times 10^{13}$ cm), and $E$ ($0.34\pm0.10$~foe \textit{vs.} $0.40\pm0.12$ foe). 
The inferred parameters are fully consistent with low-energy explosions of medium-mass red supergiant stars, in the range $12.8 - 14$ M$_{\odot}$.
With no deep pre-explosion images available for either of these two SNe, the direct detection of their progenitors was not possible. 
However, the nebular spectra obtained for both SNe were used to constrain the progenitors' mass. Following \cite{fransson:1989}, the luminosity ratio between the [Ca~II] $\lambda\lambda$ 7392, 7324 and [O~I] $\lambda\lambda$ 6300, 6364 doublets measured at late epochs ($\Re \approx 3.7$, for both SNe) favours red supergiants of moderate mass, between $12-15$~M$_{\odot}$. The same progenitor mass range is obtained using the emissivity ratio of [Ni~II] $\lambda$ 7378 to [Fe~II] $\lambda$ 7155 \citep{jerkstrand:2015}. Instead,
comparison to models of nebular spectra calculated by \cite{jerkstrand:2012} for SN~2004et would indicate a lower mass ($\lesssim 12$ M$_{\odot}$) and an higher mass ($\gtrsim 15$ M$_{\odot}$) progenitor for SN~2013am and SN~2013K, respectively, i.e. outside the mass range favoured by hydrodynamical modelling. However, we stress that specific models for low-velocity, Ni-poor Type IIP SNe are still required (Jerkstrand et al., in preparation). 

The physical properties of the progenitors of SNe~2013K and 2013am, as obtained through hydrodynamical modelling, are compared with the ejected M($^{56}$Ni) and the plateau luminosity at 50~d (L$_{50}$) in Fig.~\ref{distribution}. When compared with the sample of well-studied SNe collected in \cite{pumo:2017}, it It appears that the two events bridge the underluminous tail of Type~IIP SNe to typical, standard-luminosity events. While the total explosion energy of SN~2013am is similar to that of normal Type~IIP SN~1999em \citep[in the range $0.5 - 1$ foe, see][ and references within]{elmhamdi:2003}, SN~2013K has a lower explosion energy, albiet still comparable to normal Type IIP events \citep[for example SN~2013ab, with 0.35 foe, see ][]{bose:2015}. \cite{pumo:2017} suggests that the main parameter controlling where a SN lies within the heterogeneous Type IIP SN class, from underluminous to standard events, is the ratio $E/M_{ej}$. SNe~2013K and 2013am form a monotonic sequence with the other SNe in Fig.~\ref{distribution}, indicating, once again, the presence of a continuous distribution from faint, low-velocity, Ni-poor events to bright, high-velocity, Ni-rich objects.

\section{Acknowledgments}

This work is based (in part) on observations collected at the European Organisation for Astronomical Research in the Southern Hemisphere, Chile as part of PESSTO, (the Public ESO Spectroscopic Survey for Transient Objects Survey) ESO program 188.D-3003, 191.D-0935, 197.D-1075.

This paper is also based on observations collected at: the Copernico 1.82m and Schmidt 67/92 Telescopes operated by INAF Osservatorio Astronomico di Padova at Asiago, Italy; the Galileo 1.22m Telescope operated by Department of Physics and Astronomy of the University of Padova at Asiago, Italy; the Nordic Optical Telescope, operated by The Nordic Optical Telescope Scientific Association at the Observatorio del Roque de los Muchachos, La Palma, Spain, of the Instituto de Astrofisica de Canarias; the Liverpool Telescope operated on the island of La Palma, Spain, by Liverpool John Moores University in the Spanish Observatorio del Roque de los Muchachos of the Instituto de Astrofisica de Canarias with financial support from the UK Science and Technology Facilities Council; the Gran Telescopio Canarias (GTC), installed in the Spanish Observatorio del Roque de los Muchachos in the island of La Palma of the Instituto de Astrofisica de Canarias, Spain; 
the 3.6m Italian Telescopio Nazionale Galileo (TNG) operated by the Fundaci\'on Galileo Galilei - INAF on the island of La Palma, Spain; the Las Cumbres Observatories (LCO) global network\footnote{https://lco.global/observatory/sites/}; the  NTT 3.6m, Trappist and REM Telescopes operated by European Southern Observatory (ESO) in Chile; the SMARTS and Prompt Telescopes operated by Cerro Tololo Inter-American Observatory (CTIO) in Chile; the Australian National University 2.3-m telescope (ANU) at Siding Spring Observatory in northern New South Wales, Australia.

L.T., S.B., A.P., M.T. are partially supported by the PRIN-INAF 2014 project {\it Transient Universe: unveiling new types of stellar explosions with PESSTO}. 
M.F. is supported by a Royal Society - Science Foundation Ireland University Research Fellowship.
K.M. acknowledges support from the STFC through an Ernest Rutherford Fellowship. 
S.J.S. acknowledges funding from the European Research Council under the European Union's Seventh Framework Programme (FP7/2007-2013)/ERC Grant agreement n$^{\rm o}$ [291222]
and STFC grants ST/I001123/1 and ST/L000709/1. 
L.G. was supported by the US National Science Foundation under Grant AST-1311862. 
G.P. is supported by Ministry of Economy, Development, and Tourism's Millennium Science Initiative through grant IC120009, awarded to The Millennium Institute of Astrophysics.
C.P.G. acknowledges support from EU/FP7-ERC grant No. [615929]. 
C.B. gratefully acknowledge the support from the Wenner-Grenn Foundation.
F.E.B. acknowledges support from CONICYT-Chile (Basal-CATA PFB-06/2007, FONDECYT Regular 1141218), the Ministry of Economy, Development, and Tourism's Millennium Science
Initiative through grant IC120009, awarded to The Millennium Institute of Astrophysics, MAS.
T-W.C. acknowledges the support through the Sofia Kovalevskaja Award to P. Schady from the Alexander von Humboldt Foundation of Germany. 
We would like to thank CNTAC for the allocation of REM time through proposals CN2013A-FT-12.

We thank Andrea Melandri for useful comments on the manuscript. We also thank the anonymous referee for the thorough review of the paper.

We are grateful to {\it Istituto Nazionale di Fisica Nucleare Laboratori Nazionali del Sud} for the use of computer facilities.

The work made use of Swift/UVOT data reduced by P. J. Brown and released in the Swift Optical/Ultraviolet Supernova Archive (SOUSA). SOUSA is supported by NASA's Astrophysics Data Analysis Program through grant NNX13AF35G.
We acknowledge the Weizmann interactive supernova data repository (http://wiserep.weizmann.ac.il).

\begin{table*}
\caption{Magnitudes for the local sequence stars in the fields of SNe 2013K and 2013am.}\label{local}
\begin{tabular}{cccccccc}
\hline \\
{\bf SN 2013K}  &&&&&&&\\
\hline \\
ID & R.A. & Dec. & $U$&$B$ & $V$ & $R$ & $I$ \\
a&17:37:24.229 &$-$85:16:01.56 &$-$&17.24 (0.01)& 16.56 (0.01)& 16.27 (0.02)& 15.64 (0.05)\\ 
b&17:37:44.118 &$-$85:16:39.05  &$-$&16.62 (0.01)& 15.92 (0.01)& 15.59 (0.01)& 15.03 (0.03)\\ 
c&17:40:03.075 &$-$85:17:04.67  &$-$&16.49 (0.01)& 15.85 (0.01)& 15.49 (0.01)& 15.09 (0.04)\\ 
d&17:41:20.834 &$-$85:17:11.66  &$-$&15.34 (0.01)& 14.66 (0.01)& 14.30 (0.01)& 13.81 (0.02)\\ 
e&17:38:00.727 &$-$85:18:49.69  &$-$&17.27 (0.01)& 16.50 (0.01)& 16.13 (0.01)& 15.67 (0.02)\\ 
f&17:41:32.752 &$-$85:19:31.73  & $-$&18.03 (0.01)& 17.35 (0.01)& 16.98 (0.02)& 16.47 (0.03)\\ 
g&17:38:11.826 &$-$85:19:57.82  &$-$&18.63 (0.01)& 17.82 (0.01)& 17.35 (0.01)& 16.78 (0.02)\\ 
h&17:41:00.886 &$-$85:20:40.59  &$-$&18.34 (0.03)& 17.66 (0.01)& 17.32 (0.01)& 16.78 (0.02)\\ 
i&17:37:42.199 &$-$85:20:36.12  &$-$&18.72 (0.02)& 18.07 (0.01)& 17.76 (0.02)& 17.23 (0.04)\\ 
j&17:41:31.852 &$-$85:15:43.37  &$-$&14.41 (0.01)& 13.41 (0.01)& 12.91 (0.03)& 12.35 (0.06)\\ 
k&17:38:12.844 &$-$85:15:34.54  &$-$&18.79 (0.01)& 18.09 (0.01)& 17.73 (0.02)& 17.21 (0.02)\\ 
l&17:41:03.387 &$-$85:16:15.56  & $-$&17.42 (0.01)& 16.61 (0.01)& 16.21 (0.01)& 15.70 (0.03)\\ 
m&17:37:57.276 &$-$85:17:08.56  &$-$&18.17 (0.02)& 17.45 (0.01)& 17.08 (0.01)& 16.58 (0.02)\\ 
n&17:41:54.244 &$-$85:18:18.32  &$-$&18.77 (0.01)& 18.10 (0.02)& 17.73 (0.01)& 17.18 (0.03)\\ 
o&17:37:22.590 &$-$85:18:13.42  & $-$&18.33 (0.01)& 17.47 (0.01)& 17.05 (0.01)& 16.46 (0.02)\\ 
p&17:41:24.477 &$-$85:19:16.19  & $-$&18.01 (0.01)& 17.22 (0.02)& 16.79 (0.02)& 16.12 (0.04)\\ 
q&17:38:38.263 &$-$85:19:52.86  &$-$ &17.73 (0.01)& 16.99 (0.01)& 16.59 (0.01)& 16.07 (0.01)\\ 
r&17:41:32.707 &$-$85:21:02.86  &$-$&15.99 (0.02)& 15.20 (0.01)& 14.80 (0.02)& 14.14 (0.02)\\ 
s&17:38:30.267 &$-$85:20:59.03  &$-$&16.69 (0.03)& 16.20 (0.03)& 15.86 (0.03)& 15.35 (0.04)\\ 
t&17:41:39.802 &$-$85:17:31.23  &$-$&17.71 (0.01)& 16.89 (0.01)& 16.43 (0.01)& 15.81 (0.02)\\ 
u&17:40:44.322 &$-$85:17:49.86  &$-$&18.36 (0.01)& 17.13 (0.01)& 16.37 (0.01)& 15.61 (0.03)\\ 
v&17:36:49.463 &$-$85:17:09.99  &$-$&17.38 (0.01)& 16.65 (0.01)& 16.30 (0.01)& 15.82 (0.02)\\ 
w&17:36:36.302 &$-$85:18:35.50  &$-$&17.25 (0.02)& 16.22 (0.02)& 15.69 (0.01)& 15.04 (0.03)\\ 
x& 17:40:12.352 &$-$85:15:46.66  & $-$&13.90 (0.01)& 12.37 (0.01)& 11.52 (0.01)&  $-$\\
\hline \\
{\bf SN 2013am} &&&&&&&\\
\hline \\
ID & R.A. & Dec. & $U$&$B$ & $V$ & $R$ & $I$ \\
a&11:18:53.136 &13:05:05.12&  18.56 (0.05) & 17.46 (0.06)&  16.27 (0.04) & 15.59  (0.05)&  14.97  (0.09)\\ 
b&11:18:57.773 &13:05:30.36&  15.89 (0.03) & 15.45 (0.05)&  14.59 (0.04) & 14.09  (0.04)&  13.66  (0.06) \\
c&11:19:05.728 &13:03:28.38&  18.19 (0.03) & 18.21 (0.04)&  17.55 (0.03) & 17.16  (0.03)&  16.76  (0.06) \\
d&11:19:00.156 &13:05:34.53&  19.50 (0.08) & 19.60 (0.04)&  19.03 (0.04) & 18.69  (0.04)&  18.33  (0.06) \\
e&11:18:52.405 &13:03:35.20&  12.62 (0.03) & 14.44 (0.08)&  12.79 (0.04) & 11.99  (0.02)&  11.71  (0.04) \\
f&11:18:58.230 &13:05:54.53&  22.12 (0.83) & 21.10 (0.11)&  19.53 (0.10) & 18.30  (0.17)&  16.24  (0.34) \\
g&11:19:00.421 &13:02:21.35&  20.11 (0.13) & 18.91 (0.07)&  17.54 (0.04) & 16.75  (0.05)&  16.06  (0.10) \\
h&11:18:55.426 &13:04:48.63&  20.63 (0.28) & 19.81 (0.06)&  18.84 (0.06) & 18.31  (0.05)&  17.93  (0.06) \\
i&11:18:55.756 &13:04:04.68&  17.15 (0.04) & 18.27 (0.04)&  17.96 (0.04) & 17.88  (0.06)&  18.08  (0.09) \\
\hline\\
\end{tabular}
\end{table*}

\begin{table*}
\caption{SN 2013K: Optical photometry in the {\em UBVRI} bands. Errors are given in parentheses.}\label{phot1}
\begin{tabular}{cccccccc}
\hline \\ 
Date & MJD & $U$ & $B$ & $V$  & $R$ & $I$ & Instrument\\
\hline \\ 
20130120 & 56312.41 & $-$ & $-$  &$-$&  17.70  & $-$  & CBET3391  \\ 
20130122 & 56314.32 & $-$ &  $-$ &  17.69 (0.01) &$-$   &$-$   & EFOSC2  \\ 
20130123 & 56315.25 & $-$ &  $-$ &  17.69 (0.16) &  17.53 (0.30) &  17.25 (0.27) & PROMPT \\ 
20130124 & 56316.29 &$-$  &  18.30 (0.21) &$-$&  $-$ & $-$  & kb77  \\ 
20130204 & 56327.36 &$-$  &  $-$ &  $-$ &  17.38 (0.11) &  16.91 (0.10) & kb77  \\ 
20130209 & 56332.36 & $-$ &  18.88 (0.11) &  $-$ &  $-$ &  16.83 (0.43) & kb77  \\ 
20130211 & 56334.25 &$-$  &  18.82 (0.06) &  17.70 (0.04) &  17.21 (0.04) &  16.85 (0.09) & ANDICAM-CCD  \\ 
20130211 & 56334.35 &  $-$&  18.84 (0.07) &  17.72 (0.05) &  $-$ &  16.94 (0.07) & kb77  \\ 
20130212 & 56335.36 & $-$ &  $-$ &$-$&  17.34 (0.06) &  16.98 (0.10) & kb77  \\ 
20130212 & 56335.36 &$-$  &   $-$&  17.73 (0.05) &  $-$ & $-$  & TRAPPISTCAM  \\ 
20130213 & 56336.25 & $-$ &  18.86 (0.05) &  17.73 (0.04) &  17.24 (0.05) &  16.98 (0.07) & ANDICAM-CCD  \\ 
20130214 & 56337.36 & $-$ &  18.84 (0.07) &  $-$ &  $-$ &  16.97 (0.07) & kb77  \\ 
20130215 & 56338.36 & $-$ &  18.98 (0.08) &  17.66 (0.06) &  $-$ &  16.95 (0.08) & kb77  \\ 
20130216 & 56339.25 & $-$ &  18.94 (0.03) &  17.69 (0.08) &  17.27 (0.03) &  16.85 (0.08) & ANDICAM-CCD  \\ 
20130216 & 56339.36 &$-$&  $-$ &  $-$ &  17.35 (0.05) &  16.98 (0.09) & kb77  \\ 
20130217 & 56340.36 & $-$ &  18.97 (0.08) &  17.76 (0.06) &  17.37 (0.04) &  16.73 (0.08) & kb77  \\ 
20130217 & 56340.38 & $-$ &  18.93 (0.03) &  17.72 (0.03) &  17.28 (0.04) &  16.88 (0.19) & ANDICAM-CCD  \\ 
20130218 & 56341.36 & $-$ &  18.86 (0.10) &  $-$ &  17.19 (0.06) &  16.83 (0.09) & kb77  \\ 
20130219 & 56342.38 & $-$ &  19.01 (0.03) &  17.65 (0.02) &  17.19 (0.04) &  16.91 (0.05) & ANDICAM-CCD  \\ 
20130220 & 56343.36 & $-$ &  18.99 (0.09) &  17.75 (0.04) &  17.29 (0.04) &  16.74 (0.08) & kb77  \\ 
20130221 & 56344.25 & $-$ &  18.92 (0.03) &  17.68 (0.06) &  17.16 (0.04) &  16.74 (0.08) & ANDICAM-CCD  \\ 
20130221 & 56344.36 & $-$ &$-$&  $-$ &  $-$ &  16.84 (0.12) & kb77  \\ 
20130222 & 56345.36 & $-$ &  $-$ &  17.73 (0.05) &  17.32 (0.06) &  16.65 (0.07) & kb77  \\ 
20130223 & 56346.36 & $-$ &  19.04 (0.16) &  17.66 (0.06) &  $-$ &  16.83 (0.11) & kb77  \\ 
20130224 & 56347.36 &  $-$&  19.12 (0.12) &  $-$ &  $-$ &  16.91 (0.09) & kb77  \\ 
20130225 & 56348.36 & $-$ &  $-$ &  $-$ &  17.29 (0.09) &  16.80 (0.09) & kb77  \\ 
20130226 & 56349.36 &$-$&  19.17 (0.13) &  $-$ &  17.22 (0.11) &  $-$ & kb77  \\ 
20130227 & 56350.25 & $-$ &  19.02 (0.11) &  17.72 (0.05) &  17.09 (0.07) &  16.83 (0.08) & ANDICAM-CCD  \\ 
20130228 & 56351.36 & $-$ &  $-$ &  17.80 (0.07) & $-$ &  16.80 (0.08) & kb77  \\ 
20130301 & 56352.25 & $-$ &  19.10 (0.10) &  17.79 (0.09) &  17.08 (0.04) &  16.98 (0.09) & ANDICAM-CCD  \\ 
20130301 & 56352.36 &$-$&  $-$ &  17.77 (0.06) &  17.23 (0.06) &  16.73 (0.08) & kb77  \\ 
20130302 & 56353.38 &  $-$&  $-$ &  $-$ &  17.12 (0.38) &$-$& TRAPPISTCAM  \\ 
20130303 & 56354.25 &  $-$&  19.17 (0.04) &  17.66 (0.03) &  17.05 (0.03) &  16.66 (0.04) & ANDICAM-CCD  \\ 
20130303 & 56354.36 &  $-$&$-$&  17.75 (0.07) &  $-$ &  16.61 (0.08) & kb77  \\ 
20130305 & 56356.25 &  $-$&  19.11 (0.04) &  17.69 (0.03) &  17.09 (0.04) &  16.64 (0.05) & ANDICAM-CCD  \\ 
20130305 & 56356.36 &$-$&  19.23 (0.11) &  $-$ &  $-$ &  16.82 (0.13) & kb77  \\ 
20130306 & 56357.36 & $-$ &  19.33 (0.14) &  17.66 (0.06) &  $-$&  16.73 (0.10) & kb77  \\ 
20130310 & 56361.25 & $-$ &  19.26 (0.03) &  17.70 (0.04) &  17.10 (0.08) &  16.69 (0.11) & ANDICAM-CCD  \\ 
20130311 & 56362.25 & $-$ &  19.18 (0.04) &  $-$ &  17.10 (0.04) &  16.59 (0.04) & ANDICAM-CCD  \\ 
20130311 & 56362.31 & 20.89 (0.07) &  19.31 (0.02) &  17.79 (0.03) &$-$&$-$& EFOSC2  \\ 
20130314 & 56365.25 &  $-$&  19.36 (0.08) &  17.70 (0.09) &  17.07 (0.11) &  16.56 (0.21) & ANDICAM-CCD  \\ 
20130314 & 56365.36 &  $-$&  19.21 (0.15) &  17.78 (0.07) &  17.11 (0.10) &  16.67 (0.14) & kb73  \\ 
20130315 & 56366.36 &  $-$&  19.25 (0.11) &  $-$ &  17.10 (0.05) &  16.59 (0.08) & kb73  \\ 
20130316 & 56367.36 &  $-$&  19.45 (0.21) &  $-$ &$-$&  $-$ & kb73  \\ 
20130317 & 56368.32 &  $-$&  19.25 (0.12) &  17.74 (0.07) &  16.95 (0.08) &  16.50 (0.08) & PROMPT  \\ 
20130317 & 56368.36 &  $-$&  19.25 (0.13) &  17.71 (0.06) &  16.99 (0.04) &  16.52 (0.07) & kb73  \\ 
20130318 & 56369.25 &  $-$&  19.25 (0.03) &  17.67 (0.06) &  16.95 (0.06) &  16.51 (0.11) & ANDICAM-CCD  \\ 
20130318 & 56369.36 &  $-$&  19.16 (0.08) &  17.70 (0.08) &  $-$ &  16.54 (0.08) & kb73  \\ 
20130319 & 56370.21 & 20.83 (0.10) &  19.26 (0.03) &  17.70 (0.04) & $-$  & $-$  & EFOSC2  \\ 
20130319 & 56370.36 &  $-$&  19.18 (0.08) &  $-$ &  17.04 (0.06) &  16.43 (0.10) & kb73  \\ 
20130324 & 56375.25 &  $-$&  19.21 (0.03) &  17.75 (0.09) &  16.97 (0.03) &  16.58 (0.04) & ANDICAM-CCD  \\ 
20130326 & 56377.25 &  $-$&  19.36 (0.06) &  17.74 (0.09) &  17.02 (0.08) &  16.51 (0.12) & ANDICAM-CCD  \\ 
\hline \\ 
\end{tabular}
\end{table*}

\begin{table*}
\contcaption{SN 2013K: Optical photometry.}
\begin{tabular}{llllllll}
\hline \\ 
Date & MJD & $U$ & $B$ & $V$ & $R$  & $I$ & Instrument \\ 
\hline \\ 
20130402 & 56384.42 & $-$ &  19.58 (0.05) &  17.79 (0.03) & $-$  & $-$  & EFOSC2  \\ 
20130406 & 56388.25 &$-$&  19.59 (0.04) &  17.76 (0.04) &  16.98 (0.02) &  16.41 (0.02) & ANDICAM-CCD  \\ 
20130406 & 56388.29 & 21.57 (0.14) &  19.57 (0.04) &  17.88 (0.04) & $-$  & $-$  & EFOSC2  \\ 
20130412 & 56394.39 & $-$ & $-$  &  17.86 (0.60) &  $-$ &  $-$ & TRAPPISTCAM  \\ 
20130413 & 56395.36 &$-$&  19.60 (0.04) &  17.85 (0.04) &  17.02 (0.04) &  16.49 (0.09) & ANDICAM-CCD  \\ 
20130415 & 56397.25 & $-$ &  19.64 (0.04) &  17.89 (0.05) &  17.07 (0.06) &  16.42 (0.03) & ANDICAM-CCD  \\ 
20130418 & 56400.37 & $-$ &  19.74 (0.11) &  17.91 (0.05) &  17.11 (0.03) &  16.39 (0.06) & ANDICAM-CCD  \\ 
20130422 & 56404.35 & $-$ &  19.70 (0.05) &  18.10 (0.05) &  17.12 (0.04) &  16.51 (0.05) & ANDICAM-CCD  \\ 
20130423 & 56405.38 &  $-$& $-$  &  18.10 (0.13) &  $-$ &  $-$ & TRAPPISTCAM  \\ 
20130426 & 56408.32 & $-$ & $-$  & $-$  &  17.26 (0.36) &  $-$ & ANDICAM-CCD  \\ 
20130507 & 56419.33 & $-$ &  20.08 (0.15) &  18.29 (0.08) & $-$  & $-$  & kb77  \\ 
20130516 & 56428.26 & $-$ & $-$  &  18.88 (0.20) & $-$  & $-$  & kb77  \\ 
20130519 & 56431.26 &$-$ & $-$  &  19.50 (0.23) & $-$  &  $-$ & kb77  \\ 
20130603 & 56446.22 & $-$ & $-$  &  20.66 (0.37) &  $-$ & $-$  & kb77  \\ 
20130607 & 56450.22 & $-$ &  21.58 (0.23) &  20.13 (0.16) & $-$  & $-$  & kb77  \\ 
20130804 & 56508.10 & $-$ & $-$  &  21.05 (0.31) &  20.30 (0.27) & $-$  & EFOSC2  \\ 
20130805 & 56509.11 & $-$ & $-$  &  21.17 (0.11) &  20.31 (0.27) & $-$  & EFOSC2 \\ 
20130816 & 56520.07 & $-$ & $-$  &  21.18 (0.10) &  20.40 (0.05) & $-$  & EFOSC2  \\ 
20130913 & 56548.01 & $-$ &$-$&  21.41 (0.26) &$-$&  19.65 (0.19) & EFOSC2  \\ 
20131004 & 56569.02 &  $-$& $-$  &  21.50 (0.30) &  20.78 (0.30) &$-$& EFOSC2  \\ 
20131027 & 56592.05 &$-$&  $-$ &  21.18 (0.12) &  20.26 (0.42) &$-$& EFOSC2  \\ 
20131104 & 56600.03 & $-$ &$-$&  21.98 (0.25) &  21.03 (0.11) &$-$& EFOSC2  \\ 
\hline \\ 
\end{tabular}
\end{table*}

\begin{table*}
\caption{SN 2013K: Infrared photometry}\label{phot2}
\begin{tabular}{cccccc}
\hline \\ 
Date & MJD & $J$ & $H$ & $K$  & Instrument\\ 
\hline \\ 
20130129 & 56321.39 & 16.87 (0.12) &  16.60 (0.12) &  $-$ & SOFI  \\
20130130 & 56322.18 & $-$ &  16.63 (0.20) &  16.13 (0.16) & SOFI  \\ 
20130208 & 56331.25 & 16.65 (0.12) &  16.62 (0.13) &  16.36 (0.15) & SOFI  \\ 
20130213 & 56336.34 & 16.61 (0.23) & $-$  &  $-$ & ANDICAM-IR  \\ 
20130216 & 56339.36 & 16.43 (0.14) &  16.13 (0.24) & $-$  & ANDICAM-IR  \\ 
20130217 & 56340.38 & 16.41 (0.26) &  16.46 (0.21) &$-$& ANDICAM-IR  \\ 
20130219 & 56342.38 & 16.43 (0.16) &  16.16 (0.09) & $-$  & ANDICAM-IR  \\ 
20130221 & 56344.31 & $-$ &  16.03 (0.11) &  $-$ & ANDICAM-IR  \\ 
20130222 & 56345.26 & 16.52 (0.12) &  16.20 (0.12) &  16.04 (0.12) & SOFI  \\ 
20130227 & 56350.36 & 16.28 (0.13) &  16.13 (0.18) &  $-$ & ANDICAM-IR  \\ 
20130301 & 56352.35 &  $-$ &  16.12 (0.07) &  $-$ & ANDICAM-IR  \\ 
20130303 & 56354.35 & 16.27 (0.16) &  $-$ &   $-$& ANDICAM-IR  \\ 
20130305 & 56356.31 & 16.29 (0.11) &  15.94 (0.07) &$-$& ANDICAM-IR  \\ 
20130306 & 56357.27 & 16.36 (0.10) &  16.16 (0.10) &  15.96 (0.12) & SOFI  \\ 
20130310 & 56361.32 & 16.18 (0.09) &  $-$ & $-$  & ANDICAM-IR  \\ 
20130313 & 56364.11 & 16.23 (0.15) &  15.97 (0.12) &  15.84 (0.11) & SOFI  \\ 
20130319 & 56370.10 & 16.19 (0.14) &  15.98 (0.11) &  15.86 (0.12) & SOFI  \\ 
20130324 & 56375.32 &$-$&  15.92 (0.06) &$-$& ANDICAM-IR  \\ 
20130326 & 56377.33 & 15.88 (0.20) &  15.79 (0.13) &  $-$ & ANDICAM-IR  \\ 
20130404 & 56386.23 & 16.15 (0.10) &  15.91 (0.12) &  15.79 (0.09) & SOFI  \\ 
20130412 & 56394.23 & 16.15 (0.12) &  16.03 (0.12) &  15.83 (0.09) & SOFI  \\ 
20130418 & 56400.24 & 16.21 (0.10) &  16.01 (0.11) &  15.86 (0.13) & SOFI  \\ 
\hline \\ 
\end{tabular}
\end{table*}

\begin{table*}
\caption{SN 2013K: Optical photometry in the {\em gri} bands. The magnitude system is as for SDSS DR12, that is $SDSS = AB - 0.02$ mag. Errors are given in parentheses.}\label{phot3}
\begin{tabular}{cccccc}
\hline \\ 
Date & MJD & $g$ & $r$ & $i$  & Instrument\\ 
\hline \\
20130507 & 56419.35   &  19.27 (0.17) &  $-$ & $-$ &     kb77\\
20130512 & 56424.35   &  19.37 (0.13) & 18.21 (0.13)  & 17.54 (0.09) &   kb77\\
20130516 & 56428.27   &  19.99 (0.08)  & 18.15 (0.05) & 17.75 (0.05)  &   kb77\\
20130519 & 56431.27   &      $-$                &  18.82 (0.56) & 18.16 (0.06)     &  kb77\\
20130603 & 56446.23   &  20.86 (0.18)  & 20.06 (0.13) & 19.17 (0.14)     &kb77\\
20130607 & 56450.24   &       $-$             & 19.70 (0.21)   &$-$ & kb77 \\
\hline \\ 
\end{tabular}
\end{table*}

\begin{table*}
\caption{SN 2013am: Optical photometry in the {\em UBVRI} bands. Errors are given in parentheses.}\label{phot4}
\begin{tabular}{cccccccc}
\hline \\
Date & MJD & $U$ & $B$ & $V$ & $R$  & $I$ & Instrument\\
\hline \\
20130321 & 56372.64 & $-$ &  $-$ &$-$&  15.60 (0.30) &$-$& CBET3440  \\
20130322 & 56373.80 &$-$&  17.10 (0.09) &  16.45 (0.04) &  16.01 (0.04) &  15.65 (0.05) & AFOSC  \\
20130324 & 56375.22 &$-$&  17.06 (0.06) &  16.35 (0.06) &$-$&$-$& PROMPT  \\
20130326 & 56377.13 &$-$&  17.08 (0.05) &  16.30 (0.05) &  15.76 (0.06) &  15.34 (0.04) & PROMPT  \\
20130326 & 56377.24 &$-$&  17.06 (0.05) &  16.34 (0.05) &$-$&$-$& kb74  \\
20130328 & 56379.16 &$-$&  17.06 (0.06) &  16.42 (0.08) &  15.62 (0.07) &  15.31 (0.06) & PROMPT  \\
20130328 & 56379.24 &$-$&  17.09 (0.05) &  16.34 (0.03) &$-$&$-$& kb74  \\
20130329 & 56380.10 &$-$&  17.17 (0.06) &  16.35 (0.05) &  15.83 (0.06) &  15.31 (0.04) & PROMPT  \\
20130329 & 56380.45 &$-$&  17.08 (0.05) &  16.34 (0.04) &$-$&$-$& fs01  \\
20130330 & 56381.26 &$-$&  17.10 (0.07) &  16.28 (0.04) &$-$&$-$& fs02  \\
20130401 & 56383.24 &$-$&  17.20 (0.05) &  16.29 (0.03) &$-$&$-$& kb74  \\
20130402 & 56384.06 &$-$&  17.36 (0.04) &  16.59 (0.03) &  15.89 (0.04) &  15.37 (0.04) & TRAPPISTCAM  \\
20130403 & 56385.17 &$-$&  17.42 (0.08) &  16.32 (0.05) &  15.70 (0.07) &  15.22 (0.03) & PROMPT  \\
20130404 & 56386.24 &$-$&  17.32 (0.06) &  16.30 (0.02) &$-$&$-$& kb74  \\
20130405 & 56387.24 &$-$&  17.36 (0.07) &  16.31 (0.04) &$-$&$-$& kb74  \\
20130407 & 56389.24 &$-$&  17.55 (0.06) &  16.44 (0.07) &$-$&$-$& kb74  \\
20130411 & 56393.17 &$-$&  $-$ &  16.66 (0.18) &$-$&$-$& kb74  \\
20130413 & 56395.17 &$-$&  18.03 (0.09) &  16.57 (0.05) &$-$&$-$& kb74  \\
20130413 & 56395.19 & 18.91 (0.21) &  17.99 (0.23) &$-$&$-$&$-$& EFOSC2  \\
20130414 & 56396.17 &$-$&  17.93 (0.05) &  16.55 (0.05) &$-$&$-$& kb74  \\
20130415 & 56397.17 &$-$&  17.91 (0.07) &  16.51 (0.06) &  15.81 (0.05) &  15.21 (0.05) & PROMPT  \\
20130415 & 56397.17 &$-$&  17.94 (0.07) &  16.51 (0.04) &$-$&$-$& kb74  \\
20130419 & 56400.97 &$-$&  18.05 (0.20) &  16.55 (0.05) &  15.96 (0.16) &  15.19 (0.17) & SBIG  \\
20130419 & 56401.17 &$-$&  18.20 (0.08) &  16.63 (0.04) &$-$&$-$& kb74  \\
20130420 & 56402.17 &$-$&  18.13 (0.06) &  16.53 (0.04) &$-$&$-$& kb74  \\
20130421 & 56403.17 &$-$&  18.15 (0.09) &  16.51 (0.03) &$-$&$-$& kb74  \\
20130423 & 56405.17 &$-$&  18.17 (0.07) &  16.58 (0.04) &$-$&$-$& kb74  \\
20130423 & 56405.92 &$-$&  18.29 (0.09) &  16.46 (0.16) &  15.71 (0.23) &  15.10 (0.14) & SBIG  \\
20130425 & 56407.92 &$-$&  18.26 (0.12) &  16.56 (0.09) &  15.73 (0.63) &  15.10 (0.20) & SBIG  \\
20130426 & 56408.15 &$-$&  18.17 (0.09) &  16.47 (0.04) &$-$&$-$& kb74  \\
20130427 & 56409.15 &$-$&  18.25 (0.10) &  16.59 (0.05) &$-$&$-$& kb74  \\
20130428 & 56410.15 &$-$&  18.20 (0.08) &  16.53 (0.04) &$-$&$-$& kb74  \\
20130429 & 56411.15 &$-$&  18.29 (0.07) &  16.55 (0.04) &  $-$ &$-$& kb74  \\
20130430 & 56412.15 &$-$&  18.23 (0.07) &  16.59 (0.04) &$-$&$-$& kb74  \\
20130430 & 56412.91 &$-$&  18.30 (0.09) &  16.56 (0.18) &  15.81 (0.25) &  15.04 (0.12) & SBIG  \\
20130503 & 56415.17 &$-$&$-$&  16.74 (0.15) &$-$&$-$& kb74  \\
20130504 & 56416.97 &$-$&  18.45 (0.07) &  16.54 (0.05) &  15.72 (0.06) &  15.05 (0.08) & AFOSC  \\
20130506 & 56418.17 &$-$&  18.41 (0.09) &  16.65 (0.07) &$-$&$-$& kb74  \\
20130508 & 56420.16 &$-$&  18.58 (0.27) &$-$&$-$&$-$& kb74  \\
20130510 & 56422.19 &$-$&  18.72 (0.28) &$-$&$-$&$-$& kb74  \\
20130513 & 56425.17 &$-$&  18.75 (0.10) &  16.69 (0.07) &   &   & kb74  \\
20130513 & 56425.95 &$-$&  18.57 (0.08) &  16.53 (0.15) &  15.73 (0.25) &  15.02 (0.20) & SBIG  \\
20130515 & 56427.17 &$-$&  18.87 (0.17) &  16.76 (0.06) &  $-$ &$-$& kb74  \\
20130516 & 56428.16 &$-$&  18.87 (0.25) &$-$&$-$&$-$& kb74  \\
20130521 & 56433.02 &$-$&  18.81 (0.20) &  16.70 (0.06) &  15.82 (0.08) &  15.05 (0.06) & PROMPT  \\
20130602 & 56445.01 &$-$&  18.85 (0.13) &  16.75 (0.04) &  15.79 (0.06) &  15.05 (0.04) & PROMPT  \\
20130605 & 56448.96 &$-$&$-$&  16.86 (0.22) &$-$&$-$& PROMPT  \\
20130610 & 56453.94 &$-$&$-$&  16.96 (0.20) &$-$&$-$& PROMPT  \\
20130611 & 56454.96 &$-$&$-$& $-$  &  16.01 (0.22) &  15.13 (0.25) & PROMPT  \\
20130614 & 56457.98 &$-$&$-$&  16.83 (0.06) &  15.84 (0.05) &  15.13 (0.06) & PROMPT  \\
20130617 & 56460.96 &$-$&  19.11 (0.43) &  16.93 (0.30) &  16.15 (0.13) &  15.27 (0.07) & PROMPT  \\
20130623 & 56466.96 &$-$&  19.09 (0.06) &  16.93 (0.06) &  16.36 (0.06) &  15.13 (0.07) & LRS  \\
20130707 & 56480.40 &$-$&$-$& 18.30 (0.03)  & 17.22 (0.13) & 16.41 (0.09)  & LJT$^a$ \\
20131205 & 56631.18 &$-$&$-$&  20.50 (0.24) &  19.36 (0.25) &  18.19 (0.13) & AFOSC  \\
20131205 & 56631.18 &$-$&  $> 19.3$ &$-$&$-$&$-$& AFOSC  \\
20131210 & 56636.30 &$-$&$-$&  20.60 (0.08) &$-$&$-$& EFOSC2  \\
20140220 & 56708.32 &$-$&$-$&  21.00 (0.10) &  19.92 (0.04) &  18.81 (0.06) & EFOSC2  \\
20140223 & 56711.10 &$-$&$-$&  21.18 (0.09) &$-$&$-$& EFOSC2  \\ 
20140223 & 56711.10 &$-$&  $> 20.5$  &$-$&$-$&$-$& EFOSC2  \\ 
\hline \\ 
\end{tabular}

$^a$ = data from Zhang et al. 2014, LJT  = 2.4m Telescope, Li-Jiang Observatory
\end{table*}

\begin{table*}
\caption{SN 2013am: Infrared photometry.}\label{phot5}
\begin{tabular}{cccccc}
\hline \\ 
Date & MJD & $J$ & $H$ & $K$  & Instrument\\
\hline \\ 
20130324 & 56375.11 & 14.97 (0.05) &  14.71 (0.06) &  14.51 (0.24) & REMIR  \\
20130325 & 56376.13 & 14.88 (0.06) &  14.62 (0.18) &  14.38 (0.10) & REMIR  \\
20130329 & 56380.23 & 14.81 (0.19) &  14.43 (0.22) &  14.06 (0.24) & REMIR  \\
20130404 & 56386.18 & 14.44 (0.31) &  14.47 (0.27) &  14.22 (0.22) & SOFI  \\ 
20130412 & 56394.11 & 14.39 (0.26) &  14.19 (0.31) &  13.96 (0.25) & SOFI  \\ 
20130418 & 56400.13 & 14.25 (0.24) &  14.17 (0.26) &  13.97 (0.28) & SOFI  \\ 
20130419 & 56401.05 & 14.38 (0.09) &  14.03 (0.28) &  13.82 (0.07) & REMIR  \\
20130422 & 56404.11 & 14.38 (0.36) &  13.96 (0.26) &  13.96 (0.18) & REMIR  \\
20130425 & 56408.00 & 14.43 (0.27) &  13.95 (0.04) & $-$  & REMIR  \\ 
20130426 & 56408.00 & $-$ &  14.07 (0.23) &  13.81 (0.08) & REMIR  \\ 
20131217 & 56643.22 & 18.05 (0.34) &  17.53 (0.05) &  17.30 (0.10) & NOTCam  \\
20140213 & 56701.11 & 18.71 (0.03) &  18.04 (0.05) &  17.89 (0.05) & NOTCam  \\
20140315 & 56731.04 & 19.42 (0.14) &  18.41 (0.35) &  18.04 (0.28) & NOTCam  \\
\hline \\ 
\end{tabular}
\end{table*} 

\begin{table*}
\caption{SN 2013am: Optical photometry in the {\em ugriz} bands. The magnitude system is as for SDSS DR12, that is $SDSS = AB - 0.02$ mag. Errors are given in parentheses.}\label{phot6}
\begin{tabular}{cccccccc}
\hline \\ 
Date & MJD & $u$ & $g$ & $r$ & $i$ & $z$  & Instrument\\ 
\hline \\
20130323 &56374.90 &$-$& 16.83 (0.04)& 15.99 (0.02)& 15.87 (0.03)& 15.82 (0.03)&  RATCam\\
 20130324 &56375.88 &$-$& 16.75 (0.02)& 16.12 (0.01)& 15.83 (0.03)& 15.78 (0.03)&  RATCam\\
 20130325 &56376.88 &$-$& 16.76 (0.02)& 16.01 (0.02)& 15.85 (0.03)& 15.73 (0.02)&  RATCam\\
 20130327 &56378.85 &$-$&$-$& 15.90 (0.02)& 15.75 (0.02)& 15.64 (0.07)&  RATCam\\
 20130328 &56379.94 & 17.65 (0.02)& 16.80 (0.01)& 15.97 (0.01)& 15.78 (0.03)& 15.64 (0.01)&  RATCam\\
 20130329 &56380.99 & 17.81 (0.02)& 16.71 (0.01)& 15.99 (0.02)& 15.76 (0.02)& 15.61 (0.02)&  RATCam\\
 20130330 &56381.95 & 17.92 (0.02)& 16.80 (0.03)& 15.98 (0.01)& 15.75 (0.03)& 15.58 (0.03)&  RATCam\\
 20130401 &56383.93 & 18.14 (0.04)& 16.80 (0.02)& 15.93 (0.02)& 15.67 (0.02)& 15.53 (0.04)&  RATCam\\
 20130406 &56388.86 & 19.05 (0.07)& 16.91 (0.05)& 15.95 (0.03)& 15.71 (0.03)& 15.58 (0.02)&  RATCam\\
 20130408 &56390.90 & 19.54 (0.05)& 17.14 (0.03)& 16.05 (0.05)& 15.73 (0.03)& 15.60 (0.03)&  RATCam\\
 20130410 &56392.91 & 19.95 (0.08)& 17.20 (0.10)& 16.23 (0.11)& 15.85 (0.04)& 15.51 (0.03)&  RATCam\\
 20130413 &56395.96 & 20.22 (0.09)& 17.28 (0.08)& 16.06 (0.03)& 15.75 (0.01)& 15.56 (0.03)&  RATCam\\
 20130415 &56397.98 & 20.29 (0.12)& 17.33 (0.02)& 16.12 (0.01)& 15.84 (0.03)& 15.54 (0.01)&  RATCam\\
 20130417 &56399.90 & 21.17 (0.32)& 17.24 (0.04)& 16.18 (0.03)& 15.70 (0.02)& 15.52 (0.01)&  RATCam\\
 20130421 &56403.91 &$-$& 17.31 (0.04)& 16.08 (0.02)& 15.70 (0.02)& 15.39 (0.03)&  RATCam\\
 20130423 &56405.90 &$-$&$-$& 16.19 (0.05)& 15.76 (0.05)& 15.51 (0.02)&  RATCam\\
 20130427 &56409.00 &$-$& 17.32 (0.05)&$-$& 15.66 (0.09)& 15.34 (0.24)&  RATCam\\
 20130430 &56412.02 &     $-$        & 17.37 (0.04)& 16.02 (0.02)& 15.65 (0.02)&   $-$          &  RATCam\\
 20130506 &56418.01 &$-$& 17.48 (0.05)& 16.06 (0.02)& 15.60 (0.05)& 15.26 (0.01)&  RATCam\\
 20130512 &56424.91 &$-$& 17.45 (0.09)& 16.00 (0.02)& 15.61 (0.03)&$-$&  RATCam\\
 20130517 &56429.93 &$-$& 17.75 (0.03)& 16.03 (0.01)& 15.55 (0.04)& 15.34 (0.01)&  RATCam\\
 20130521 &56433.91 &$-$& 17.77 (0.05)& 16.03 (0.02)& 15.55 (0.03)& 15.32 (0.02)&  RATCam\\
 20130611 &56454.87 &$-$& 18.12 (0.03)& 16.80 (0.09)& 15.97 (0.04)& 15.37 (0.01)&  RATCam\\
 20130328 &56379.25 &$-$& 16.67 (0.03)& 15.92 (0.04)&$-$&$-$&  kb74 \\
 20130401 &56383.24 &$-$& 16.72 (0.03)& 15.90 (0.03)& 15.69 (0.02)&$-$&  kb74 \\
 20130404 &56386.24 &$-$& 16.86 (0.03)& 15.87 (0.03)&$-$&$-$&  kb74 \\
 20130405 &56387.24 &$-$& 16.86 (0.06)& 15.95 (0.06)& 15.56 (0.06)&$-$&  kb74 \\
 20130407 &56389.24 &$-$& 16.99 (0.04)& 15.99 (0.07)&$-$&$-$&  kb74 \\
 20130411 &56393.17 &$-$& 17.16 (0.10)& 16.03 (0.10)& 15.75 (0.10)&$-$&  kb74 \\
 20130413 &56395.17 &$-$& 17.21 (0.05)& 16.10 (0.05)& 15.83 (0.05)&$-$&  kb74 \\
 20130414 &56396.17 &$-$& 17.22 (0.05)& 16.14 (0.05)&$-$&$-$&  kb74 \\
 20130419 &56401.17 &$-$& 17.37 (0.03)& 16.10 (0.03)& 15.68 (0.04)&$-$&  kb74 \\
 20130420 &56402.17 &$-$& 17.32 (0.04)& 16.08 (0.04)& 15.69 (0.03)&$-$&  kb74 \\
 20130421 &56403.17 &$-$& 17.25 (0.04)& 15.99 (0.03)& 15.63 (0.05)&$-$&  kb74 \\
 20130423 &56405.17 &$-$& 17.28 (0.05)& 16.02 (0.04)& 15.62 (0.04)&$-$&  kb74 \\
 20130426 &56408.16 &$-$& 17.35 (0.06)& 16.05 (0.04)& 15.62 (0.06)&$-$&  kb74 \\
 20130427 &56409.16 &$-$& 17.34 (0.05)& 16.03 (0.04)& 15.53 (0.07)&$-$&  kb74 \\
 20130428 &56410.16 &$-$& 17.36 (0.06)& 16.06 (0.04)& 15.58 (0.05)&$-$&  kb74 \\
 20130429 &56411.16 &$-$& 17.30 (0.04)& 15.99 (0.03)& 15.53 (0.05)&$-$&  kb74 \\
 20130430 &56412.16 &$-$& 17.38 (0.04)& 16.02 (0.03)& 15.57 (0.05)&$-$&  kb74 \\
 20130503 &56415.17 &$-$& 17.43 (0.11)& 16.17 (0.10)& 15.71 (0.05)&$-$&  kb74 \\
 20130506 &56418.18 &$-$&$-$& 16.11 (0.06)& 15.65 (0.07)&$-$&  kb74 \\
 20130508 &56420.18 &$-$&$-$& 16.07 (0.26)&$-$&$-$&  kb74 \\
 20130513 &56425.18 &$-$& 17.61 (0.07)& 16.06 (0.06)& 15.64 (0.05)&$-$&  kb74 \\
 20130515 &56427.17 &$-$& 17.70 (0.06)& 16.11 (0.05)& 15.65 (0.04)&$-$&  kb74 \\
 20130329 &56380.40 &$-$& 16.62 (0.05)& 15.92 (0.03)& 15.76 (0.04)& 15.63 (0.06)&  fs02 \\
 20130329 &56380.55 &$-$& 16.75 (0.04)& 15.93 (0.02)& 15.78 (0.03)& 15.58 (0.05)&  fs01 \\
 20130330 &56381.27 &$-$& 16.72 (0.04)& 15.94 (0.02)& 15.73 (0.02)& 15.59 (0.05)&  fs02\\
\hline \\ 
\end{tabular}
\end{table*}

\begin{table*}
\caption{SN 2013K: Journal of spectroscopic observations. The phase is relative to the estimated date of explosion reported in Table~1.} \label{telescope_spec}
\begin{tabular}{c c c c c c}
\hline
Date	  &MJD      &Phase	& Instrumental & Range & Resolution        \\
&     &[d]         & configuration$^1$& [\AA]  & [\AA]		 \\ 
\hline
20130122 &  56314.49       & 12.5    &NTT+EFOSC2+gr13 & 3650$-$9250& 18\\ 
20130127 &  56319.54       & 17.5   &NTT+EFOSC2+gr11+gr16 &3350$-$10 000& 14\\
20130130 &  56322.43       & 20.4   &NTT+EFOSC2+gr11+gr16 & 3350$-$10 000& 13\\
20130206 &  56329.45       & 27.4   & NTT+EFOSC2+gr11+gr16 & 3350$-$10 000& 13\\
20130220 &  56343.45       & 41.4   &NTT+EFOSC2+gr13 &3650$-$9250 & 17\\ 
20130303 &  56354.51      &  52.5   &NTT+EFOSC2+gr13 &3650$-$9250 & 17\\ 
20130310 &  56361.49       & 59.5   &NTT+EFOSC2+gr13 & 3650$-$9250& 18\\ 
20130318 &  56369.48       & 67.5   &NTT+EFOSC2+gr13 & 3650$-$9250& 18\\ 
20130402 &  56384.50       & 82.5   &NTT+EFOSC2+gr13 & 3650$-$9250& 18\\ 
20130406 &  56388.47       & 86.5   &NTT+EFOSC2+gr13 & 3650$-$9250& 18\\ 
20130414 &  56396.48       & 94.5   &NTT+EFOSC2+gr13 & 3650$-$9250& 17\\ 
20130418 &  56400.42       & 98.4   &NTT+SOFI+BG & 9347$-$16440& 25\\ 
20130420 &  56402.50       & 100.5   &NTT+EFOSC2+gr13 & 3650$-$9250& 18\\ 
20130508 &  56419.58       & 117.6 &ANU2.3+WiFeS+R3000        & 5400$-$9225 &  2 \\
20131103 &  56600.21       &  298.2 &NTT+EFOSC2+gr13 & 3650$-$9250& 18\\ 
\hline
\end{tabular}

NTT = 3.6m New Technology Telescope, ESO (La Silla, Chile); ANU2.3 = Australian National University 2.3m Telescope (Siding Spring Observatory, Australia). {\sl Phase} (column 3) is relative to the estimated explosion date, MJD = 56302.0.  {\sl Resolution} (column 6)  is estimated from the FWHM of night sky lines.
\end{table*}

\begin{table*}
\caption{SN 2013am: Journal of spectroscopic observations. The phase is relative to the estimated date of explosion reported in Table~1.} \label{telescope_spec2}
\begin{tabular}{c c c c c c}
\hline
Date	  &MJD      &Phase	& Instrumental & Range & Resolution        \\
&     &[d]         & configuration$^1$& [\AA]  & [\AA]		 \\ 
\hline
20130322 &  56373.84       & 2.3     &Ekar+AFOSC+gm4 & 3400$-$9000& 24   \\       
20130324 &  56375.81       & 4.3     &ANU2.3+WiFeS+B3000+R3000       &  3500$-$7600 &2\\
20130327 &  56379.11       & 7.6     &ANU2.3+WiFeS+B3000+R3000         & 3500$-$7600& 2 \\
20130328 &  56379.34       & 7.8     &FTN+FLOYDS          &3200$-$10000 & 13    \\
20130329 &  56380.35       & 8.8   &FTN+FLOYDS          &3200$-$10000 & 13\\
20130330 &  56381.31       & 9.8   & FTN+FLOYDS          &3200$-$10000 & 13\\
20130331 &  56382.10       & 10.6   &TNG+LRS+LR-B+LR-R  &3200$-$10000 &10    \\
20130402 &  56384.33       & 12.8   &NTT+EFOSC2+gr13 &3650$-$9250& 18\\
20130403 &  56385.35       & 13.8    & NTT+SOFI+BG+RG   &9350$-$25000 &25 \\
20130405 &  56387.33       & 15.8   &   NTT+EFOSC2+gr13 &3650$-$9250& 18\\
20130406 &  56388.33       & 16.8   & FTN+FLOYDS          &3200$-$10000 & 13   \\
20130407 &  56389.34       & 17.8  &FTN+FLOYDS          &3200$-$10000 & 13    \\
20130409 &  56390.80       & 19.3   &ANU2.3+WiFeS+B3000+R3000       &  3500$-$7600 &2 \\
20130410 &  56392.90       & 21.4   &TNG+LRS+LR-B+LR-R  &3200$-$10000 &10     \\
20130411 &  56393.32       & 21.8   &   FTN+FLOYDS          &3200$-$10000 & 13    \\
20130411 &  56393.38       & 21.9   &NTT+SOFI+BG+RG &9350$-$25000 &25 \\
20130412 &  56394.34       & 22.8   &NTT+EFOSC2+gr13 &3650$-$9250& 18 \\
20130417 &  56399.35       & 27.8   &NTT+SOFI+BG+RG &9350$-$25000 &25 \\
20130418 &  56400.34       & 28.8   &NTT+EFOSC2+gr13 & 3650$-$9250& 18\\
20130423 &  56405.88       & 34.4   &Pennar+BC+300tr/mm   &3400$-$7800 &10 \\
20130425 &  56407.83       & 36.3   &Pennar+BC +300tr/mm   &3400$-$7800 &10\\
20130501 &  56413.96       & 42.5   &TNG+LRS+LR-B     &3200$-$8000& 10   \\
20130504 &  56416.94       & 45.4   &Ekar+AFOSC+gm4 & 3500$-$8200 &24   \\ 
20130513 &  56425.94       & 54.4  &Ekar+AFOSC+gm4+VPH6 & 3500$-$9000& 24 \\ 
20130623 &  56466.91        & 95.4  &TNG+LRS+LR-B+LR-R  &3200$-$10000 &10    \\
20130629 &  56472.85       & 101.3&Ekar+AFOSC+VPH6 & 4500$-$9300&24   \\ 
20131226 &  56652.45       & 280.9 & NTT+EFOSC2+gr13 & 3650$-$9250& 18  \\
20140204 &  56692.22       & 320.7\\ 
\hline
\end{tabular}

Ekar = Copernico 1.82m Telescope, INAF (Mt. Ekar, Asiago, Italy);  Pennar = Galileo 1.22m Telescope, DFA University of Padova (Asiago, Italy); NTT = New Technology Telescope 3.6m, ESO (La Silla, Chile); WHT = William Herschel Telescope 4.2m ; TNG = 3.6m Telescopio Nazionale Galileo, INAF (La Palma, Spain); FTN =  Faulkes Telescope North 2.0m Telescope, Las Cumbres Observatory LCO (Haleakala, Hawai, USA); ANU2.3 = Australian National University 2.3m Telescope (Siding Spring Observatory, Australia); GTC =  10.4 m Gran Telescopio Canarias (La Palma, Spain). {\sl Phase} (column 3) is relative to the estimated explosion date, MJD = 56371.5. {\sl Resolution} (column 6)  is estimated from the FWHM of night sky lines.
\end{table*}

\label{lastpage}
\end{document}